\documentclass[useAMS,usenatbib]{mn2e}

\newcommand{\dev}{\ensuremath{\,\mathrm{d}}}

\usepackage{amssymb}
\usepackage{amsmath}
\usepackage[dvips]{graphicx}
\usepackage{epsfig}
\usepackage{multicol}

\date{Accepted ... Received ...; in original form ...}
\title{Gamma-ray burst afterglows from trans-relativistic blast wave simulations}
\author[Van Eerten et al.]{H.J. van Eerten$^{1}$\thanks{E-mail: H.J.vanEerten@uva.nl}, K. Leventis$^{^1}$, Z. Meliani$^2$, R.A.M.J. Wijers$^1$, R. Keppens$^{2,3,4}$\\
$^{1}$Astronomical Institute 'Anton Pannekoek', PO box 94248, 1090 SJ Amsterdam, the Netherlands\\
$^{2}$Centre for Plasma Astrophysics, K.U. Leuven, Celestijnenlaan 200B, 3001 Leuven, Belgium\\
$^{3}$FOM-Institute for Plasma Physics Rijnhuizen, Nieuwegein, The Netherlands\\
$^{4}$Astronomical Institute, Utrecht University, The Netherlands}

\begin{document}

\maketitle

\begin{abstract}
We present a study of the intermediate regime between ultra-relativistic and nonrelativistic flow for gamma-ray burst afterglows. The hydrodynamics of spherically symmetric blast waves is numerically calculated using the \textsc{amrvac} adaptive mesh refinement code. Spectra and light curves are calculated using a separate radiation code that, for the first time, links a parametrisation of the microphysics of shock acceleration, synchrotron self-absorption and electron cooling to a high-performance hydrodynamics simulation. For the dynamics we find that the transition to the nonrelativistic regime generally occurs later than expected, that the Sedov-Taylor solution overpredicts the late time blast wave radius and that the analytical formula for the blast wave velocity from \cite{Huang1999} overpredicts the late time velocity by a factor 4/3. Also we find that the lab frame density directly behind the shock front divided by the fluid Lorentz factor squared remains very close to four times the unshocked density, while the effective adiabatic index of the shock changes from relativistic to nonrelativistic. For the radiation we find that the flux may differ up to an order of magnitude depending on the equation of state that is used for the fluid and that the counterjet leads to a clear rebrightening at late times for hard-edged jets. Simulating GRB030329 using predictions for its physical parameters from the literature leads to spectra and light curves that may differ significantly from the actual data, emphasizing the need for very accurate modelling. Predicted light curves at low radio frequencies for a hard-edged jet model of GRB030329 with opening angle 22 degrees show typically two distinct peaks, due to the combined effect of jet break, non relativistic break and counterjet. Spatially resolved afterglow images show a ring-like structure.
\end{abstract}

\section{Introduction}

Gamma-ray burst (GRB) afterglows can be explained from the interaction between an initially relativistic shock wave of hot fluid and the medium surrounding the burster. On passage of the shock electrons get accelerated to relativistic velocities (even with respect to the already relativistic local fluid flow) and small scale magnetic fields are generated. Under influence of the magnetic field, the electrons will produce synchrotron radiation, which will be seen by the observer. This model has been very successful when applied to broadband afterglow data, but thus far model predictions have been made using simplifying assumptions for the blast wave structure (approximating the blast wave width by a homogeneous slab, e.g. \citealt{Wijers1997, Meszaros1997, Sari1998, Rhoads1999}) or from analytical solutions in either the ultrarelativistic or the nonrelativistic regime (e.g. \citealt{Granot1999, Gruzinov1999, Wijers1999, Frail2000}).

Since the beginning of the decade, fluid simulations have been performed to study afterglow blast waves and their resulting spectra (see \citealt{Granot2001, Downes2002}). More recent simulations have been used to address the specific theoretical issue of the visible effect of the blast wave encountering a density perturbation \citep{Nakar2007, vanEerten2009b}. Very recently \citet{Zhang2009} studied the transition to the transrelativistic regime and the spreading of a collimated outflow, using an adaptive mesh technique for the fluid simulation. They made some simplifying assumptions for the radiation mechanism, when compared to the early analytical efforts (e.g. \citealt{Granot1999}), such as approximating the cooling time by the lab frame time and ignoring synchrotron self-absorption.

The aim of this paper is to present a theoretical and qualitative study of the transition regime between relativistic and nonrelativistic blast waves and the effect on the light curves and spectra at various wavelengths, using adaptive mesh relativistic fluid simulations for blast waves from an explosion in a homogeneous medium, while including \emph{all} details of the synchrotron radiation mechanism that have been used for earlier analytical estimates. Also we present resolved afterglow images. We study spherical blast waves and sharp edged jets obtained by taking conic sections from a spherically symmetric fluid flow.

Obviously these simulations do not yet fully address the complete GRB afterglow picture of a realistic, 2D-dynamical jet, which we address in future work. However, some GRB afterglows have power law decays that last for months without a jet break, and thus may be (nearly) spherical. These are of course already addressed in the present work. Also, by studying the conic sections from spherical flows, we already address some aspects of jet behaviour, which allows us to probe some outstanding issues, such as whether the receding jet may lead to visible features
in the late light curve, and whether a dynamical jet break must be truly achromatic. Finally, any fluid flow behaviour typical to higher-dimensional simulations, like lateral spreading of the jet, is best understood from a direct comparison to one-dimensional simulations and its effects on the light curve will in practice be modeled as a deviation from the heuristic description based on analytical approximations and one-dimensional simulations (i.e. as an additional smooth jet break). A companion paper is in preparation that will discuss the practical consequences for broadband afterglow data fitting from the underlying model from this paper.

This paper is organized as follows. In section \ref{radiation_code_section} we discuss our radiation code and how it expands upon an approach outlined earlier in \cite{vanEerten2009}, hereafter EW09. A proper treatment of synchrotron radiation and shock wave generation of accelerated particles and small scale magnetic fields requires us to trace some additional quantities along with the fluid quantities.

In section \ref{fluid_dynamics_section} we provide the details of our simulations that assume typical GRB parameters. We show how the blast wave starts out in the ultrarelativistic regime and smoothly approaches the nonrelativistic regime. We discuss the consequences of different equations of state for the fluid and how our simulations differ from analytical approximations for the nonrelativistic regime. We show how the fluid lab frame density divided by the fluid Lorentz factor squared right behind the shock remains \emph{always} close to four times that in front of the shock, even though we have differing adiabatic indices in both the relativistic and nonrelativistic regimes. Three additional quantities needed to be traced and we present results for the behaviour of these three: the accelerated electron number density, the magnetic field energy density and the accelerated particle distribution upper cut-off Lorentz factor. We explain how calculation of the latter especially is numerically challenging and how it shapes the spectrum beyond the cooling break.

In section \ref{spectra_section} we take our results from section \ref{fluid_dynamics_section} and calculate spectra and light curves. We calculate spectra at 1, 10, 100, 1,000 and 10,000 days in observer time. We separately discuss the different factors contributing to the shape of the light curves: the equation of state, the evolution of the magnetic field and the evolution of the accelerated particle distribution.

We then turn to the specific case of GRB 0303029 in section \ref{GRB030329_section}. We take the explosion parameters that have been established for this burst by previous authors to set up a simulation. We qualitatively compare the resulting light curves to radio data at different wavelengths, assuming both a spherical explosion and a hard edged jet with opening angle of $22$ degrees. We provide spatially resolved radio images and make a qualitative prediction for the expected signal at radio wavelengths that will be observable with the next generation of telescopes, like \textsc{lofar}.

We discuss our results in \ref{discussion_section}. In the appendices we provide additional technical details on the numerical implementation of our approach and a discussion on the theoretical limitations and assumptions of our approach.

\section{The radiation code}
\label{radiation_code_section}
In this paper we follow the approach first outlined in EW09, where we calculate spectra and light curves from the output of a relativistic hydrodynamics (RHD) code using a separate radiation code. For the RHD simulations we use \textsc{amrvac}, a high performance code that includes adaptive-mesh refinement (AMR) (see \citealt{Keppens2003, Meliani2007}). \textsc{amrvac} calculates the evolution of the following conserved variables:
\begin{equation}
D = \gamma \rho', \quad \vec{S} = \gamma^2 h' \vec{v}, \quad \tau = \gamma^2 h' - p' - \gamma \rho' c^2,
\end{equation}
with $\gamma$ the Lorentz factor, $\rho'$ the proper density, $h'$ the relativistic (i.e. including rest mass) enthalpy density, $\vec{v}$ the three velocity, $p'$ the pressure and $c$ the speed of light. In the entire paper, all comoving quantities will be primed.

In the second stage we use a radiation code to obtain the received flux for a given observer frequency, time and distance, from the local values of conserved variables at any contributing point in the fluid (we also use two auxiliary quantities, $\gamma$ and $p'$, that \textsc{amrvac} stores as well in order to facilitate its calculation of the time evolution of the conserved variables). The radiation mechanism that is considered is synchrotron radiation and a number of parameters have been introduced in EW09 that capture the underlying radiation and shock microphysics. There are four of these `ignorance' parameters. The fraction of the thermal energy that resides in the tangled-up magnetic field that is generated by the passage of a shock $\epsilon_\textrm{B}$ usually has a value around $0.01$. The fraction of electrons $\xi_\textrm{N}$ that is accelerated into a relativistic power law distribution in energy also by the passage of a shock is usually of order unity in the relativistic regime. The thermal energy fraction captured by these electrons $\epsilon_\textrm{E} \backsim 0.1$ and minus the slope of the electron distribution $p \backsim 2.5$.

The flux calculated by the radiation code is given by
\begin{equation}
F_\nu = \frac{1+z}{r_\textrm{obs}^2} \int \frac{ \dev^2 P_\textrm{V} }{\dev \nu \dev \Omega} ( 1 - \beta \mu ) c \dev A \dev t_\textrm{e}.
\label{flux_equation}
\end{equation}
Here $z$ denotes redshift, $r_\textrm{obs}$ denotes the observer luminosity distance, $\dev^2 P_\textrm{V} / \dev \nu \dev \Omega$ the received power per unit volume, frequency and solid angle, $\dev A$ the \emph{equidistant surface} element given by the intersection of the fluid grid with that surface from which radiation is poised to arrive exactly at $t_\textrm{obs}$, and $t_\textrm{e}$ the emission time. Note that in this terminology, \emph{flux} is defined per unit frequency. The integral
\begin{displaymath}
\int (1 - \beta \mu) c \dev A \dev t_\textrm{e}
\end{displaymath}
is effectively an integral over the entire radiating volume. $\mu$ is the angle between the local fluid velocity and the observer position, $\beta$ the fluid velocity in units of $c$ and the factor $(1 - \beta \mu)$ is a retardation effect due to the moving of the radiating source. The detailed dependency of the received power on the ignorance parameters and local fluid conditions is explained in EW09. However, in that paper only ultra-relativistic flows were addressed and in order to include subrelativistic and nonrelativistic flows as well, a number of features were added to our radiation code. Also we have added synchrotron self-absorption and the possibility to resolve the signal from the fluid into an image on the sky. We now have a generic radiation code that is capable of calculating the spatially resolved synchrotron radiation profile from an arbitrary fluid flow. The additional physics that we have included is explained below, with some of the practical numerical issues discussed separately in appendix \ref{numerics_section}. 

\subsection{Realistic equation of state}
In EW09 we applied a fixed adiabatic index $\Gamma_\textrm{ad}$ equation of state (EOS)
\begin{equation}
p' = (\Gamma_\textrm{ad} - 1) e'_\textrm{th},
\end{equation}
where $e'_\textrm{th}$ is the thermal energy density. In practice $\Gamma_\textrm{ad}$ was always set to $4/3$. However, when following a fluid from the relativistic regime (with flow velocities $\backsim c$ and thermal energy density dominating the rest mass energy density) down to the classical regime, this fixed adiabatic index becomes too restrictive. We therefore apply a Synge-like EOS \citep{Synge1957} that results in an effective adiabatic index varying smoothly from $4/3$ to its classical limit $5/3$:
\begin{equation}
p' = \frac{\rho' c^2}{3} \left( \frac{e'}{\rho' c^2} - \frac{\rho' c^2}{e'} \right),
\end{equation}
where $e'$ denotes the comoving energy density including rest mass, $e' = \rho' c^2 + e'_\textrm{th}$. This EOS has already been applied in \textsc{amrvac} (see \citealt{Meliani2008, Meliani2004}). Also, because the radiation code reads both the conserved variables as well as $p'$ from disc directly, it does not invoke any EOS itself, and no change in the radiation code was needed. The resulting effective adiabatic index is  given by
\begin{equation}
\Gamma_\textrm{ad,eff} = \frac{5}{3} - \frac{1}{3} \left( 1 - \frac{\rho'^2 c^4}{e'^2} \right).
\end{equation}
The effect of an advanced EOS on the behaviour of the fluid is profound and we discuss this in detail in section \ref{lightcurves_section}.
\subsection{Electron cooling}
\label{electron_cooling_section}

The shape of the observed spectrum from a single fluid cell, if electron cooling does not play a role, follows directly from the dimensionless function $Q ( \nu' / \nu'_\textrm{m} )$, first introduced in EW09. It has the limiting behaviour $Q \propto (\nu' / \nu'_\textrm{m})^{1/3}$ for small $(\nu' / \nu'_\textrm{m})$ and $Q \propto (\nu' / \nu'_\textrm{m})^{(1-p)/2}$ for large $(\nu' / \nu'_\textrm{m})$. The received power depends on this shape and on the local fluid quantities via
\begin{equation}
\frac{ \dev^2 P_\textrm{V} }{\dev \nu \dev \Omega} = \frac{ (p-1) \sqrt{3} q_\textrm{e}^3}{8 \pi m_e c^2} \frac{\xi_\textrm{N} n B'}{\gamma^3 (1-\beta \mu)^3} Q \left( \frac{\nu'}{\nu'_\textrm{m}} \right).
\label{power_equation}
\end{equation}
Here $n$ denotes the lab frame number density (of \emph{all} electrons, both accelerated and thermal). $B'$ denotes the local comoving magnetic field strength, calculated from the thermal energy density after the passage of a shock. $m_\textrm{e}$ and $q_\textrm{e}$ denote electron mass and charge, respectively. The frequency $\nu'_\textrm{m}$ is the synchrotron peak frequency, and it is related to the lower cut-off Lorentz factor $\gamma'_\textrm{m}$ of the power law accelerated electrons via
\begin{equation}
 \nu'_\textrm{m} = \frac{3 q_\textrm{e}}{4 \pi m_e c} \gamma'^2_\textrm{m} B'.
\label{critical_frequency_equation}
\end{equation}
If cooling plays no role, the evolution of $\gamma'_\textrm{m}$ is completely adiabatic, which has as a consequence that the total fraction of the local thermal energy density residing in the power-law accelerated particle distribution remains fixed. $\gamma'_\textrm{m}$ will be related to $e'_\textrm{th}$ throughout the
 downstream fluid according to
\begin{equation}
 \gamma'_\textrm{m} = \left( \frac{p - 2}{p - 1} \right) \frac{ \epsilon_E e'_\text{th}}{ \xi_\textrm{N} n' m_\textrm{e} c^2}.
\label{gamma_m_from_e_th_equation}
\end{equation}

When cooling \emph{does} play a role, however, this picture is changed. It now becomes necessary to introduce an upper cut-off Lorentz factor $\gamma'_\textrm{M}$ as well. In a single fluid element, no accelerated electrons with energies above $\gamma'_\textrm{M}$ will be found, because these have cooled to energies at or below $\gamma'_\textrm{M}$. The temporal evolution of any electron Lorentz factor $\gamma'_\textrm{e}$, and therefore of $\gamma'_\textrm{m}$ and $\gamma'_\textrm{M}$ as well, is given by
\begin{equation}
 \frac{ \dev \gamma'_\textrm{e}}{ \dev t'} = \frac{ \gamma'_\textrm{e}}{3 n'} \frac{ \dev n'}{ \dev t'} - (\gamma'_{\textrm{e}})^2 \frac{ \sigma_\textrm{T} B'^2}{ 6 \pi m_\textrm{e} c},
\label{kinetic_equation}
\end{equation}
where $\sigma_\textrm{T}$ is the Thomson cross section, $m_e$ the electron mass and $t'$ the comoving time. The final term in this equation reflects synchrotron radiation losses, and if it is omitted only the adiabatic cooling term is left and it can be shown that this will result in the aforementioned fixed relation between $\gamma'_\textrm{e}$ and $e'_\textrm{th}$. In light of the previous subsection on the EOS, it may be worth noting that equation \ref{kinetic_equation} is derived for a relativistic electron distribution with adiabatic index $4/3$ and that this remains valid even if the bulk of the fluid becomes nonrelativistic. After all, the power-law accelerated electrons are relativistic by definition.  

The above has the following consequences for the simulations and radiation code. Because for low values of $\gamma'_\textrm{e}$, the radiation loss term can be neglected next to the adiabatic expansion term, we will not apply equation \ref{kinetic_equation} to $\gamma'_\textrm{m}$ and we will continue to calculate $\gamma'_\textrm{m}$ locally using equation \ref{gamma_m_from_e_th_equation} in the radiation code. For $\gamma'_\textrm{M}$ this is not an option and we numerically solve equation \ref{kinetic_equation} in \textsc{amrvac}, resetting $\gamma'_\textrm{M}$ to a high value wherever we detect the passage of a shock. This reset implements the shock-acceleration of particles. In appendix \ref{numerics_section} we discuss the numerical issues of this approach in some more detail. Also, in appendix \ref{gyral_radius_appendix} we show that the gyral radius even for the high energy electrons contributing to the observed spectrum (within the frequency range under consideration, $10^8-10^{18}$ Hz) is orders of magnitude smaller then the relevant fluid scales.

Finally, we summarize the consequences of electron cooling for the spectrum discussed earlier in EW09. The received power from a fluid element is now given by
\begin{equation}
\frac{ \dev^2 P_\textrm{V} }{\dev \nu \dev \Omega} \propto \frac{\xi_\textrm{N} n B'}{\gamma^3 (1-\beta \mu)^3} \mathcal{Q} \left( \frac{\nu'}{\nu'_\textrm{M}}, \frac{\nu'}{\nu'_\textrm{m}} \right),
\end{equation}
with the relations between the upper and lower cut-off Lorentz factors and their corresponding critical frequencies given by equation \ref{critical_frequency_equation}. $\mathcal{Q}$ is a generalisation of $Q$ and the flux at frequencies $\nu'$ above $\nu'_\textrm{M}$ drops exponentially. That the resulting spectrum from the entire fluid does not show an exponential drop is due to the fact that there will always be some fluid elements contributing for which $\nu'_\textrm{M}$ is still sufficiently high. The effect of this `hot region' close to the shock front (with a size that depends on the observer frequency) on the composite synchrotron spectrum from a shock will be a steepening of the slope by $-1/2$ instead. The cooling break is found at that frequency for which the width of the hot region becomes comparable to the width of the blast wave.

\subsection{Magnetic field energy evolution}

The magnetic field directly behind a shock has been parametrised using
\begin{equation}
\frac{B'^2}{8 \pi} = e'_B = \epsilon_B e'_\text{th}. 
\end{equation}
Furthermore we assumed the number of magnetic flux lines threading a surface comoving with a fluid element to remain invariant, resulting in
\begin{equation}
 e'_\textrm{B} \propto \rho'^{4/3}.
\label{magnetic_field_proportionality_equation}
\end{equation}
For relativistic fluids this implies that the fraction $\epsilon_B$ remains fixed downstream, because $e'_\textrm{th} \propto \rho'^{\Gamma_\textrm{ad}}$. For a changing adiabatic index it is no longer possible to calculate $e'_\textrm{B}$ a posteriori from $e'_\textrm{th}$, since the relation between the two is now no longer fixed. It becomes necessary to numerically solve in \textsc{amrvac} the equation
\begin{equation}
\frac{ \dev}{\dev t'} \frac{e'_B}{\rho'^{4/3}} = 0. 
\label{magnetic_field_equation}
\end{equation}
Like $\gamma'_\textrm{M}$, we reset $e'_B$ whenever a shock is encountered. The practical implementation of the evolving magnetic field is again discussed in appendix \ref{numerics_section}. We note here that the assumption of frozen field lines is not essential, and that we can in principle include different magnetic field behaviour either by adding a source term to equation \ref{magnetic_field_equation} (parametrising for example, magnetic field decay through reconnection) or by implementing a different equation entirely.

\subsection{Changing fraction of accelerated particles}
Although $\xi_N$, the fraction of electrons accelerated by the passage of a shock is often assumed to be of the order unity for highly relativistic blast waves, it has to be lower at late times because otherwise there would not be enough energy available per accelerated electron to create a relativistic distribution (in other words, to ensure that $\gamma'_\textrm{m} > 1$). We have implemented this change in our code by replacing user parameter $\xi_N$ by $\xi_\textrm{N,NR}$, that is, the fraction of electrons that is accelerated in the nonrelativistic limit. The fraction at the relativistic limit we set to one. Because $\gamma \beta$ is the most direct measure of how relativistic the fluid flow locally is, we have parametrised the simplest possible smooth transition between both limiting cases by
\begin{equation}
 \xi_\textrm{N} = \frac{\beta \gamma + \xi_\textrm{N,NR}}{1.0 + \beta \gamma}.
\end{equation}
Whenever the passage of a shock is detected, \textsc{amrvac} resets the number density of accelerated electrons $n'_\textrm{acc}$ according to $n'_\textrm{acc} = \xi_\textrm{N} n'$, with $\xi_N$ determined using the equation above. As with the magnetic field energy density, we now need to follow $n'_\textrm{acc}$ explicitly. Because $n'_\textrm{acc}$ is a number density, its evolution is described by a continuity equation, following
\begin{equation}
 \frac{\partial}{\partial t} n'_\textrm{acc} \gamma + \frac{\partial}{\partial x^i} n'_\textrm{acc} \gamma v^i = 0,
\label{n_acc_equation}
\end{equation}
and is therefore easily implemented in \textsc{amrvac}.

\subsection{Synchrotron self-absorption}
\label{ssa_section}

In previous work we have solved equation \ref{flux_equation} by first integrating over $A$ for a given emission time $t_\textrm{e}$ (and thus for a single snapshot), followed by an integration over $t_\textrm{e}$. If we switch the order of the integrations then the integral over $t_\textrm{e}$ represents the solution to a linear radiative transfer equation without absorption, with the intensity given by
\begin{equation}
I_\nu = \int \frac{ \dev^2 P_\textrm{V} }{\dev \nu \dev \Omega} ( 1 - \beta \mu ) c \dev t_\textrm{e}.
\end{equation}
The integral over $A$ then represents a summation over all rays. The full linear radiative transfer equation including synchrotron self-absorption has the form
\begin{equation}
 \frac{\dev I_\nu}{\dev z} = - \alpha_\nu I_\nu + j_\nu,
\end{equation}
with $j_\nu \equiv \dev^2 P_\textrm{V} / \dev \nu \dev \Omega$ and $dz \equiv c \dev t_{obs} = ( 1 - \beta \mu ) c \dev t_\textrm{e}$. The synchrotron self-absorption coefficient is given by
\begin{equation}
  {\alpha}_{\nu'}' = -\frac{1}{8\pi m_\textrm{e} {\nu'}^{2}}\int\limits_{{\gamma}_\textrm{m}'}^{{\gamma}_\textrm{M}'}
 \frac{\dev{P'}_{<e>}}{\dev\nu'}\,{\gamma_\textrm{e}'}^{2}\,\frac{\partial}{\partial \gamma_\textrm{e}'}\Big[ \frac{N_\textrm{e}'\!(\gamma_\textrm{e}')}{{\gamma_\textrm{e}'}^{2}} \Big] \,\dev\gamma'_\textrm{e}.
\label{ssa_equation}
\end{equation}
Here $\dev P'_{<e>} / \dev{\nu'}$ denotes the emitted power per \emph{ensemble} electron and $N_\textrm{e}'(\gamma'_\textrm{e})$ the electron number density for relativistic electrons accelerated to $\gamma'_\textrm{e}$. Integrating $N'_e(\gamma'_e)$ over possible electron Lorentz factors yields $n'_\textrm{acc}$ by definition. These quantities are defined and explained in detail in EW09 (see also appendix \ref{ssa_subsection}).

In this treatment of the self-absorption coefficient we only take into account transitions between already occupied energy levels of electrons, leading to the integration limits of equation \ref{ssa_equation} being exactly $\gamma'_\textrm{m}$ and $\gamma'_\textrm{M}$. In this way we ignore stimulated emission arising from a population inversion below $\gamma'_\textrm{m}$. This results in values of the absorption coefficient that are larger by a factor of $3(p+2)/4$ when compared to \cite{Granot2002}.

In our radiation code, we now calculate the linear radiative transfer equation for each individual ray by not integrating over the two-dimensional surface $A$ (to get a single flux value from the collection of rays) until after the final snapshot has been processed. In addition to allowing us to include the effect of self-absorption, we now also get a spatially resolved signal from the fluid, showing the expected ring structure (extending predictions from \citealt{Granot2001apr} to the nonrelativistic regime). We use an adaptive-mesh type approach to $A$ in order to ensure an adequate spatial resolution, see appendix \ref{adaptive_mesh_subsection}.

\section{Fluid dynamics}
\label{fluid_dynamics_section}
In this section we describe the setup of our relativistic fluid simulations and compare the results against the theoretically expected behaviour.

\subsection{Expected early and late time behaviour}
Both the early and late time behaviour of the fluid can be described by a self-similar solution that is determined completely from the explosion energy $E$ and the circumburst number density $n_0$. 

At early stages, the Blandford-McKee (BM) solution \citep{Blandford1976} for relativistic blast waves predicts the following relation between the shock front fluid Lorentz factor $\Gamma$ and the explosion time ($t$, which is the same as the emission time $t_\textrm{e}$):
\begin{equation}
 \Gamma^2 = \frac{17 E}{16 \pi \rho_0 t^3 c^5}.
\end{equation}
The density $\rho_0$ is related to the number density through the proton mass: $\rho_0 = m_\textrm{p} n_0$. The shock radius $R(t)$ is then given by
\begin{equation}
 R(t) = c t \left( 1 - \frac{1}{16 \Gamma^2} \right).
\end{equation}
To lowest order $R(t)$ is just $c t$, while the shock front fluid velocity $\beta \backsim 1$. Further analytical equations for the fluid profile (in terms of pressure $p$, Lorentz factor $\gamma$, number density $n$, etc.), behind the shock front can be found in \citealt{Blandford1976}.

At late stages the evolution of the blast wave is described by the Sedov-Taylor (ST) solution \citep{Sedov1959, Taylor1950}. For a fixed adiabatic index $5/3$, the shock radius is now given by
\begin{equation}
 R(t) \approx 1.15 \left( \frac{E t^2}{\rho_0} \right)^{1/5},
\label{Sedov_R_equation}
\end{equation}
which follows directly from dimensional analysis (except for the numerical constant). In this classical approximation, the speed of light $c$ does not appear. The shock front Lorentz factor is approximately one, while $\beta$ can be found from $\beta \equiv \dev R(t) / c \dev t$. Again analytical formulae for the fluid profile exist in the literature \citep{Sedov1959}.

At some point in time the evolution of the blast wave will no longer be adequately described by the BM solution but will become more and more dictated by the ST solution. An estimate for the turning point \citep{Piran2005} can be made by equating the explosion energy to the total rest mass energy that is swept up:
\begin{equation}
 E = \rho_0 c^2 \frac{4}{3} \pi R_\textrm{NR}^3.
\label{t_NR_equation}
\end{equation}
Solving for $R_\textrm{NR}$ returns the approximate radius at which the original explosion energy in the blast wave is no longer dominant over its rest mass energy.

Analytical estimates for the bulk fluid flow velocity including the intermediate regime also exist. One such example is found in \citet{Huang1999} and we discuss it in more detail as well as compare their prediction for $\gamma(t)$ right behind the shock front directly against our simulation results in section \ref{blastwave_velocity_section}.

\subsection{Setup of simulations}

We have performed a number of simulations using the typical values for a GRB exploding into a homogeneous medium. We set up our simulations starting from the BM solution. The isotropic explosion energy $E = 1\cdot10^{52}$ erg, the medium number density $n_0 = 1 $cm$^{-3}$. We have set the initial shock Lorentz factor to 10 (and the fluid Lorentz factor therefore $\backsim 7$, differing by a factor $\sqrt{2}$). Although both \textsc{amrvac} and the radiation code are able to deal with far higher Lorentz factors, the focus for this research is on the transition to the nonrelativistic regime and for that purpose this relatively low Lorentz factor is sufficient. We have continued the simulations until the fluid proper velocity in the lab frame $\beta \backsim 0.01$.  

We have used both the advanced equation of state and a fixed adiabatic index at $4/3$ and $5/3$. In the advanced EOS simulation we have also calculated the other quantities mentioned in the previous section: $\epsilon_\textrm{B}, \xi_\textrm{N}$ and $\gamma'_\textrm{M}$. The value at the shock front for $\epsilon_\textrm{B}$ was set to the standard 0.01 and the non-relativistic limit for $\xi_\textrm{N}$ was set at $\xi_\textrm{N,NR} = 0.1$. Sufficiently high values for $\gamma'_M$ at the shock front are chosen, generally on the order of $10^7$.

In \textsc{amrvac} it is the number of refinement levels that determines the accuracy of the simulation. We have used 17 levels of refinement and 120 cells at the lowest refinement level. The grid was initially taken to run from $10^{16}$ cm to $10^{19}$ cm. The effective spatial resolution due to adaptive mesh refinement was therefore $\backsim 1.27 \cdot 10^{12}$ cm. This should be compared against the width of the blast wave at the start of the simulation, when it is the smallest. This is approximately equal to $R(t)/\Gamma^2 \backsim 3 \cdot 10^{15}$ cm, for a starting shock Lorentz factor of 10. 

Convergence of our results has been checked by performing simulations at different refinement levels and by simulations running for a shorter time on a smaller grid (thereby increasing the resolution). For the light curves and spectra we have used simulations with a shorter running time of $12.2 \cdot 10^3$ days. At this stage the fluid velocity directly behind the shock is still six percent of light speed, but we have full coverage up to $10,000$ days in observer time. The corresponding grid size is $1 \cdot 10^{17}$ cm to $6.7 \cdot 10^{18}$ cm, leading to an effective resolution of $8.3 \cdot 10^{11}$ cm.
On a standard desktop PC\footnote{For example, an Intel dual core 1600 MHz processor with 4 GB of ram.} the RHD simulations typically took a few days to complete and the radiation calculation a few hours.

\subsection{Results}

\subsubsection{Blast wave velocity}
\label{blastwave_velocity_section}

\begin{figure}
\includegraphics[width=\columnwidth]{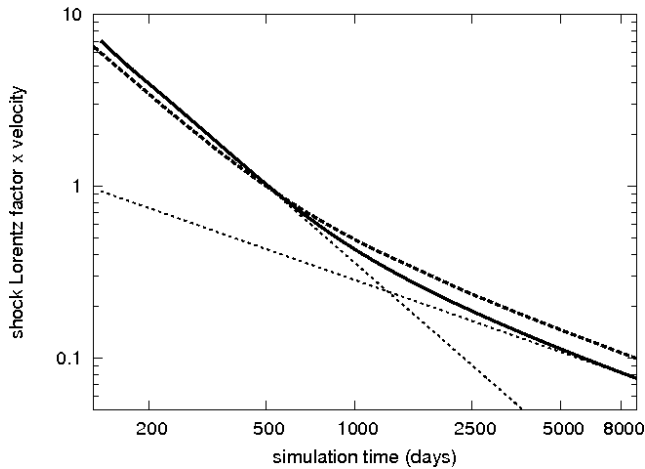}
\caption{$\beta \gamma$ at the shock front for the advanced EOS simulation (solid line), along with its asymptotic behaviour both at early and late stages. For comparison we have also plotted a prediction from \citet{Huang1999} (see text).}
\label{lfacbeta_figure}
\end{figure}
The solid line in figure \ref{lfacbeta_figure} shows $\beta \gamma$ at the shock front for the advanced EOS simulation. The expected scaling behaviour at the early stage is dictated by $\Gamma \propto t^{-3/2}$ and at the late stage by $\beta \propto t^{-3/5}$. We have plotted this asymptotic behaviour as well, setting the early stage scaling coefficient from the initial value at $\beta \gamma \backsim 7$ and the late stage scaling coefficient at $\beta \gamma \backsim 0.016$ (this point lies far to the right outside the plot). The shock velocity is shown to smoothly evolve from the BM solution to the ST solution. The meeting point of the asymptotes at $t \approx 1290$ days lies at $\beta \gamma \approx 0.244$. At this point $\beta \gamma$ for the fluid $\approx 0.33$, so the fluid is still moving at a significant fraction of the speed of light.

According to eq. \ref{t_NR_equation}, the predicted radius for the transition to occur is $R_\textrm{NR} \approx 0.38$ parsec for the initial explosion energy and circumburst density that we have used, corresponding to a lab frame time $t_\textrm{NR} \approx 450$ days. We therefore conclude that \emph{the transition point from the relativistic to the nonrelativistic regime is far later than predicted by $t_\textrm{NR}$}.

Also plotted in fig. \ref{lfacbeta_figure} is the predicted value for $\beta \gamma$ from \citet{Huang1999}, which we have implemented as follows. The starting point is
\begin{equation}
 \frac{\dev \gamma}{\dev m} = - \frac{\gamma^2 - 1}{M_\textrm{ej} + 2\gamma m},
\label{huang_equation}
\end{equation}
the differential equation proposed by the authors to depict the expansion of GRB remnants, simplified to the adiabatic case. Here $m$ denotes the rest mass of the swept-up medium and $M_\textrm{ej}$ the mass ejected from the GRB central engine. Our approach starting from the BM solution is a limiting case where $M_\textrm{ej} \downarrow 0$. The $M_\textrm{ej}$ term was included by \citet{Huang1999} to incorporate a coasting phase. When solving eq. \ref{huang_equation} we will use a very high ($\backsim 10^{7}$) initial bulk fluid Lorentz factor $\gamma_0$ and by assuming $M_\textrm{ej} \backsim E / 2\gamma_0 c^2$ we converge on the limiting scenario used in our simulations. Eq. \ref{huang_equation} can be analytically solved to yield
\begin{equation}
 ( \gamma - 1) M_\textrm{ej} c^2 + ( \gamma^2 - 1) m c^2 = E,
\end{equation}
which (numerically) leads to $\gamma(t)$ once we apply
\begin{equation}
 m = \frac{4}{3} \pi R^3 n_0 m_\textrm{p},
\end{equation}
and
\begin{equation}
 R(t) = \int_0^t \beta(\tau) c \dev \tau.
\end{equation}
Here $t$ is measured in the simulation lab frame (i.e. it does not refer to observer time).

The resulting curve for $\beta \gamma$ initially lies below the simulation result, but ends up above at $4/3$ times the simulation value. The initial and final slopes for the analytical $\beta \gamma$ curve are correct by construction. We conclude that the approach from \cite{Huang1999} initially underestimates the BM phase and significantly overestimates the late stage flow velocity. The transition point between the relativistic and nonrelativistic regime also lies at an earlier time for the analytical curve, closer to the analytically predicted $t_\textrm{NR}$.

\subsubsection{Blast wave radius}
\begin{figure}
\includegraphics[width=\columnwidth]{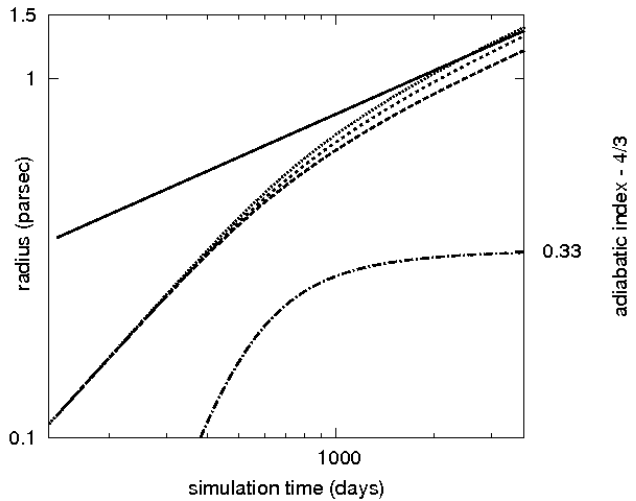}
\caption{The resulting blast wave radii as a function of lab frame time for different simulations. The steady slope line shows the radius as predicted by the ST solution. The different simulations end up in the asymptotic regime with different radii: the $\Gamma_{ad} = 5/3$ ends up above the ST solution, the advanced EOS below the ST solution between the others and $\Gamma_{ad} = 4/3$ the lowest. The bottom curve shows the effective adiabatic index for the advanced EOS, minus $4/3$. It starts at approximately zero at the left of the plot and proceeds to its asymptotic limit $1/3$ in the nonrelativistic case.}
\label{radii_figure}
\end{figure}
In figure \ref{radii_figure} we plot the blast wave radius as a function of lab frame time for three different simulations: fixed adiabatic index at $4/3$ and $5/3$ and using the advanced EOS. Also we plot the radius as predicted from eq. \ref{Sedov_R_equation} and, for the advanced EOS simulation, the difference between the effective adiabatic index and its relativistic limit $\Gamma_{ad} = 4/3$. The latter illustrates how relativistic the fluid still is in terms of temperature (as opposed to flow velocity). At the intersecting point for the $\gamma \beta$ asymptotes at 1290 days the effective adiabatic index is already quite close ($\Gamma_\textrm{ad,eff} \approx 1.63$) to its nonrelativistic limiting value. After 3800 days, when the time evolution of all the radii has become practically indistinguishable from $R(t) \propto t^{2/5}$ from the ST solution we still see a difference between the different radii. At this time the ST radius is 1.358 parsec, the $\Gamma_\textrm{ad} \equiv 4/3$ radius is 1.197 parsec, the $\Gamma_\textrm{ad} \equiv 5/3$ radius is 1.388 parsec and the advanced EOS radius 1.313 parsec. Taking the advanced EOS radius as a standard, this implies that ST overpredicts the radius at this stage by 3.4 percent, $\Gamma_\textrm{ad} \equiv 5/3$ overpredicts the radius by 5.7 percent and $\Gamma_\textrm{ad} \equiv 4/3$ underpredicts the radius by 8.9 percent. Because all radii follow close to the same temporal evolution at this stage, these errors will only very gradually become smaller throughout the further evolution of the blast wave. Therefore, a derivation of the quantity $E/\rho_0$ from the radius using the Sedov-Taylor equation \ref{Sedov_R_equation}, is likely to overpredict its value by approximately 18 percent within any time interval of practical interest.

\subsubsection{Density and energy profiles}
\begin{figure}
\includegraphics[width=\columnwidth]{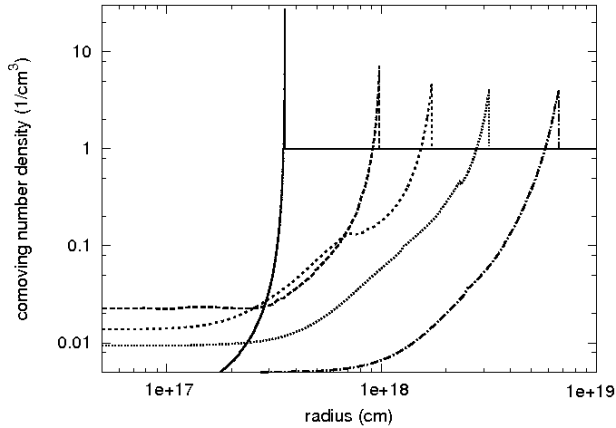}
\caption{Comoving number density profile. The profiles were taken at times corresponding to emission arrival times for the closest part of the shock front (i.e. with velocity directly towards the observer) at 10, 100, 1,000 and 10,000 days. Listed in increasing number and including the initial profile, these times correspond to lab frame times of 137, 387, 761, 2,227 and 12,583 days. Later times correspond to curves peaking further to the right in the plot.}
\label{density_figure}
\end{figure}
\begin{figure}
\includegraphics[width=\columnwidth]{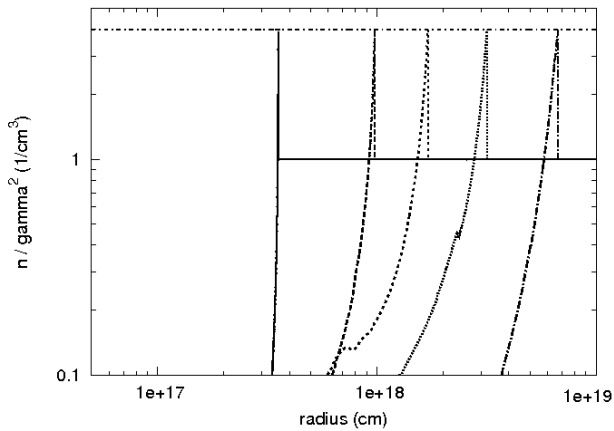}
\caption{Lab frame number density divided by $\gamma^2$. This effectively scales the shock profile to 4 at the shock front. The same lab frame times as in fig. \ref{density_figure} have been used.}
\label{ngg_figure}
\end{figure}
In figure \ref{density_figure} we have plotted the comoving fluid number density profile (of the protons \emph{or} the electrons, not both) at five different moments in time. Because later on we discuss spectra and light curves up to an observer time of 10,000 days, we have chosen emission times corresponding to arrival times of the shock front (using $t_\textrm{obs} = t - R(t)/c$) up to 10,000 days as well. The earliest fluid profile shows the initial conditions calculated from the BM solution when the shock Lorentz factor is 10. After some time, the number density at the shock front can be seen to tend to the value predicted from the shock-jump conditions for a strong classical shock, which is $n_0 ( \Gamma_\textrm{ad} + 1 )/(\Gamma_\textrm{ad} - 1) = 4 n_0$  for the classical value of the adiabatic index $5/3$. 

What is shown in figure \ref{ngg_figure} is an interesting feature of the blast wave, which is that the lab frame density directly behind the shock divided by the squared fluid Lorentz factor directly behind the shock, $D / \gamma^2 = 4 \rho_0$ \emph{throughout the entire simulation}. This can be seen analytically to hold from the Rankine-Hugoniot relations in both the ultrarelativistic and nonrelativistic case, even though the adiabatic indices have different fixed values, from
\begin{equation}
\frac{D}{\rho_0 \gamma^2} = \frac{\Gamma_\textrm{ad} + 1 / \gamma }{ \Gamma_\textrm{ad} - 1},
\end{equation}
which can be viewed as the relativistic generalization of the classical compression ratio and holds for arbitrary $\gamma$. When we use an advanced EOS, where we let $\Gamma_\textrm{ad}$ smoothly evolve from $4/3$ to $5/3$, we see from the figure that this generalized compression ratio remains very close to four even at intermediate times. We make use of this feature for the shock detection algorithm (see appendix \ref{shock_detection_appendix}).

\begin{figure}
\includegraphics[width=\columnwidth]{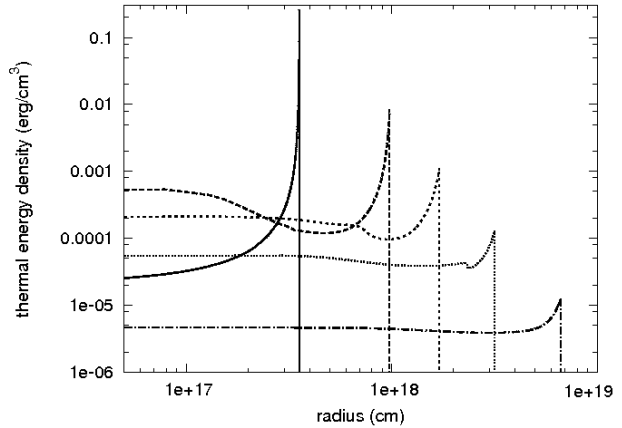}
\caption{Thermal energy density profile for the same lab frame times as in fig. \ref{density_figure}.}
\label{e_figure}
\end{figure}
In figure \ref{e_figure} we have plotted the thermal energy density at the same times as the number density. Unlike the density at the shock front, the thermal energy density is not expected to tend to a fixed value. The ST solution instead predicts a steep decline $\propto R(t)^{-3}$, which is why the final shock front thermal energy density is many orders of magnitude smaller than the initial shock front thermal energy density.

\subsubsection{Magnetic field and particle acceleration}
\begin{figure}
\includegraphics[width=\columnwidth]{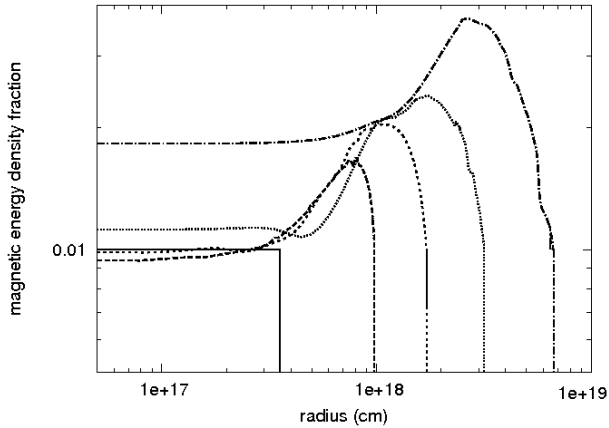}
\caption{The fraction of the thermal energy that resides in the magnetic field energy $\epsilon_\textrm{B}$ for the same lab frame times as in figure \ref{density_figure}.}
\label{epsilon_B_figure}
\end{figure}
We now turn to those quantities calculated in \textsc{amrvac} solely to aid in the construction of spectra and light curves, and that have no feedback on the dynamics. In figure \ref{epsilon_B_figure} we have plotted $\epsilon_\textrm{B}$. Because we assumed the number of field lines through a fluid surface element to remain frozen (see eqn. \ref{magnetic_field_proportionality_equation}), the magnetic field energy density declined less rapidly than the thermal energy and as a consequence the local fraction $\epsilon_\textrm{B}$ increased. A discussion on the merit of our assumption about the magnetic field behaviour is outside the scope of this work (and from particle-in-cell simulations it can certainly be argued that it is not perfect, see e.g. \citealt{Chang2008}). However, our plot does show that at least it does not lead to unphysical values or strong inconsistencies. The maximum value for $\epsilon_\textrm{B}$ found in fig. \ref{epsilon_B_figure} is 0.037 (up from 0.01 at the shock front), which is not unreasonably large and, besides, occurs far downstream in a region that will contribute negligibly to the observed flux. We emphasize that $\epsilon_B$ is a relative measure and that \emph{both} thermal and magnetic energy densities drop steeply, both with respect to earlier times and with respect to their value at the shock front at any time. The numerical method presented in this paper for parametrizing the magnetic field energy density is quite general and can be readily modified to study different parametrizations. 

\begin{figure}
\includegraphics[width=\columnwidth]{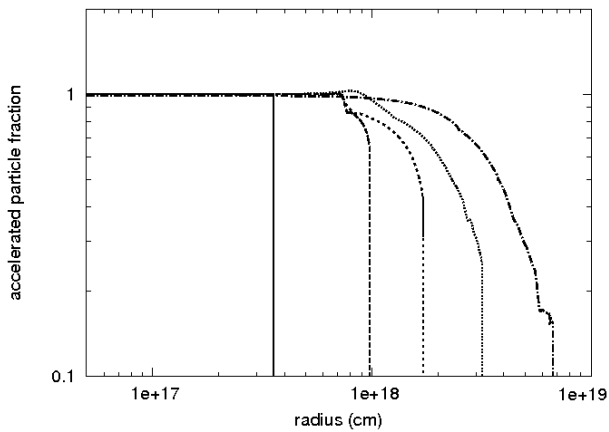}
\caption{$\xi_\textrm{N}$, the fraction of electrons accelerated to a power law distribution for the same lab frame times as in fig. \ref{density_figure}.}
\label{xi_N_figure}
\end{figure}
In figure \ref{xi_N_figure} we have plotted the fraction $\xi_\textrm{N}$ of electrons that are accelerated to a power-law distribution. This fraction was taken to smoothly decrease from unity in the relativistic regime down to 0.1 for our simulation in the nonrelativistic regime. The rightmost profile, with the shock front arriving at 10,000 days, has $\xi_\textrm{N}$ down to 0.16.

\begin{figure}
\includegraphics[width=\columnwidth]{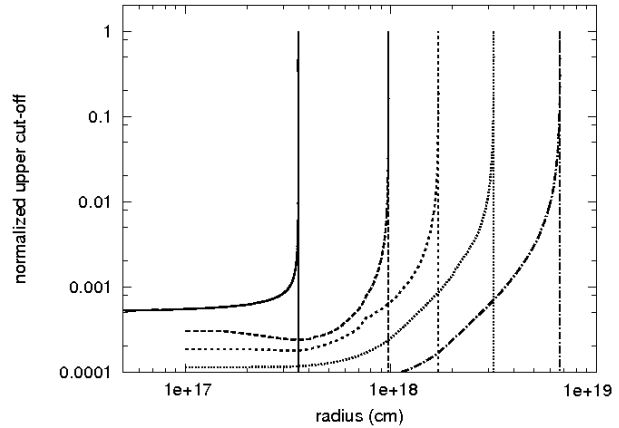}
\caption{Upper cut-off Lorentz factor of the power law distribution $\gamma'_\textrm{m}$, normalized to 1 at the shock front, for the same lab frame times as in fig. \ref{density_figure}.}
\label{lfac_M_figure}
\end{figure}
In figure \ref{lfac_M_figure} we have plotted the normalized values for $\gamma'_\textrm{M}$, the upper cut-off Lorentz factor of the power-law particle distribution. Although formally $\gamma'_\textrm{M}$ should be reset to infinity at the shock front, we picked a value corresponding to a cut-off above $10^{18}$ Hz through the entire simulation, for a fluid element heading directly towards the observer (see also section \ref{electron_cooling_section} and appendix \ref{numerics_section}). In our simulation settings, this results in $\gamma'_\textrm{M}$ peak values of the order $10^7$, and because these values are arbitrary as long as they are sufficiently large, we have normalized the $\gamma'_\textrm{M}$ profiles. The profiles show two things. First, they show the steep decline directly after the injection of new hot electrons. This steep decline and the $\backsim 7$ orders or magnitude difference between shocked and unshocked $\gamma'_\textrm{M}$ are numerically challenging, which is why we implemented the logarithm of $\gamma'_\textrm{M}$ in our code instead (again see appendix \ref{numerics_section}). Second, the width of the profile is a measure for the size of the hot region discussed in section \ref{electron_cooling_section}. It can be seen from the figure that the width of the profile increases over time. The width will nevertheless remain smaller than the width of the density profile by far. In our simulations, we resolve the $\gamma'_\textrm{M}$ profile and use it to determine the local refinement level.

\section{Spectra and light curves}
\label{spectra_section}
Using the simulation data described in the previous section we have calculated spectra and light curves at various observation times and frequencies. We have saved a total number of 10,000 snapshots of the fluid profile, with $10.8 \cdot 10^4$ seconds between consecutive snapshots, corresponding to a resolution $c \dev t \backsim 3 \cdot 10^{15}$ cm. Although this resolution is of the same order as the initial shock width, it is still sufficient at the early stage because the shock initially nearly keeps up with its own radiation. The effective resolution is given by $c \dev t / \Gamma^2 \backsim 3 \cdot 10^{13}$ cm, which is only a factor 10 larger than the spatial resolution of $10^{12}$ cm and corresponds to a temporal resolution of $\dev t \backsim 1000$ seconds. It is therefore ensured that the blast wave in the initial stage is covered by over a hundred snapshots.

\subsection{Expected spectral and temporal behaviour}

The scaling behaviour for the critical frequencies and the flux is well known from analytical estimations assuming a homogeneous radiating slab directly behind the shock front with fluid properties determined via either the BM or ST solution (see e.g. \citealt{Wijers1999, Frail2000, Granot2002}). We summarize the interstellar medium (ISM) scalings below, with $\nu_\textrm{m}$ denoting the peak frequency, $\nu_\textrm{A}$ the synchrotron self-absorption critical frequency and $\nu_\textrm{c}$ the cooling break frequency. At the observer times and frequencies in this paper we find either $\nu_\textrm{A} < \nu_\textrm{m} < \nu_\textrm{c}$ or $\nu_\textrm{m} < \nu_\textrm{A} < \nu_\textrm{c}$.

In the relativistic limit, the corresponding scalings are
\begin{eqnarray}
 \nu_\textrm{A} & \propto & \bigg\{ \begin{array}{ll} t^0, & \quad \nu_\textrm{A} < \nu_\textrm{m} \\ t^{-(3p+2)/2(p+4)}, & \quad \nu_\textrm{A} > \nu_\textrm{m} \end{array}  \\
 \nu_\textrm{m} & \propto & t^{-3/2}, \\
 \nu_\textrm{c} & \propto & t^{-1/2}, 
\end{eqnarray}
for the critical frequencies. Note that $t$ now refers to \emph{observer} times. The flux above both peak and self-absorption break scales as
\begin{equation}
F \propto \bigg\{ \begin{array}{ll} \nu^{(1-p)/2} t^{3(1-p)/4}, & \quad \nu < \nu_c \\ \nu^{-p/2} t^{(2-3p)/4}, & \quad \nu > \nu_\textrm{c} \end{array}.
\end{equation}
If $\nu_\textrm{A} < \nu_\textrm{m}$ we get the following flux scaling below the peak break:
\begin{equation}
F \propto \bigg\{ \begin{array}{ll} \nu^2 t^{1/2}, & \quad \nu < \nu_\textrm{A} \\ \nu^{1/3} t^{1/2}, & \quad \nu > \nu_\textrm{A} \end{array}.
\end{equation}
If $\nu_\textrm{A} > \nu_\textrm{m}$ we have below the self-absorption break
\begin{equation}
F \propto \bigg\{ \begin{array}{ll} \nu^2 t^{1/2}, & \quad \nu < \nu_\textrm{m} \\ \nu^{5/2} t^{5/4}, & \quad \nu > \nu_\textrm{m} \end{array}.
\end{equation}

In the nonrelativistic limit the scalings are
\begin{eqnarray}
 \nu_\textrm{A} & \propto & \bigg\{ \begin{array}{ll} t^{6/5}, & \quad \nu_\textrm{A} < \nu_\textrm{m} \\ t^{-(3p-2)/(p+4)}, & \quad \nu_\textrm{A} > \nu_\textrm{m} \end{array}  \\
 \nu_\textrm{m} & \propto & t^{-3}, \\
 \nu_\textrm{c} & \propto & t^{-1/5}, 
\end{eqnarray}
for the critical frequencies. The flux above both peak and self-absorption break scales as
\begin{equation}
F \propto \bigg\{ \begin{array}{ll} \nu^{(1-p)/2} t^{(21-15p)/10}, & \quad \nu < \nu_\textrm{c} \\ \nu^{-p/2} t^{(4-3p)/2}, & \quad \nu > \nu_\textrm{c} \end{array}.
\end{equation}
If $\nu_\textrm{A} < \nu_\textrm{m}$ we get the following flux scaling below the peak break:
\begin{equation}
F \propto \bigg\{ \begin{array}{ll} \nu^2 t^{-2/5}, & \quad \nu < \nu_\textrm{A} \\ \nu^{1/3} t^{8/5}, & \quad \nu > \nu_\textrm{A} \end{array}.
\end{equation}
If $\nu_\textrm{A} > \nu_\textrm{m}$ we have below the self-absorption break
\begin{equation}
F \propto \bigg\{ \begin{array}{ll} \nu^2 t^{13/5}, & \quad \nu < \nu_\textrm{m} \\ \nu^{5/2} t^{11/10}, & \quad \nu > \nu_\textrm{m} \end{array}.
\end{equation}

The summary above shows that only the temporal behaviour of the break frequencies and fluxes is altered by the transition to the nonrelativistic regime. We therefore do not expect spectra calculated from our simulations covering the transition to differ in slope from the slopes calculated above. The light curve slope, however, may differ.

\subsection{Spectra}
\begin{figure}
\includegraphics[width=\columnwidth]{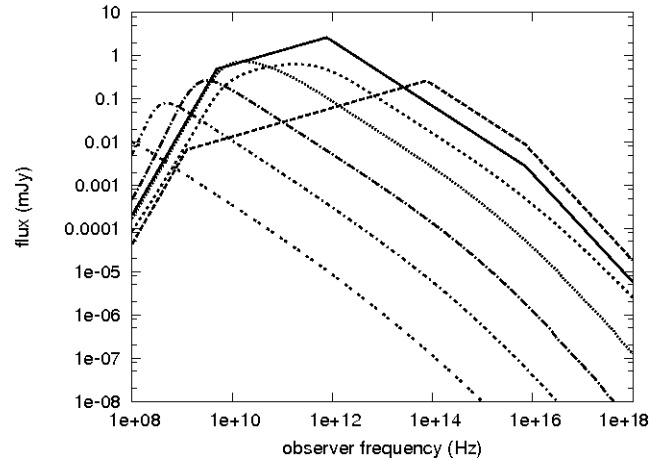}
\caption{Spectra at different observer times. The smooth curves show simulated spectra at different observer times: 1, 10, 100, 1,000 and 10,000 days, with later observed spectra having lower flux in the high frequency range. For comparison we have included predicted slopes at the different power law regimes after 1 day, for both $\xi_\textrm{N} = 1$ (solid line) and $\xi_\textrm{N} = 0.1$ (dashed line).}
\label{spectra_figure}
\end{figure}
In figure \ref{spectra_figure} we have plotted spectra for a number of different observation times, ranging from 1 day to 10,000 days. For comparison we have also plotted the different power law slopes for 1 day as predicted by \citet{Granot2002}, where we have added a dependency on $\xi_\textrm{N}$. We plot predictions for both $\xi_\textrm{N} = 1$ and $\xi_\textrm{N} = 0.1$. It can be seen that the simulated spectrum still lies closer to the $\xi_\textrm{N} = 1$ prediction, just as we would expect for an early time spectrum. Because of shifts in both flux level and position of the spectral breaks for different values of $\xi_\textrm{N}$, the flux does not always lie in between the analytical predictions. For example, because the peak frequency $\nu_\textrm{m}$ for the simulation lies close to that of the $\xi_\textrm{N} = 1$ prediction and flux lies also closer to $\xi_\textrm{N} = 1$ but below, the resulting flux at higher frequencies ends up below both predictions.

Figure \ref{spectra_figure} proves that our method works and that the asymptotic behaviour for the spectral slopes matches the predicted slopes. For frequencies above the self-absorption break and below the cooling break this merely confirms that the synchrotron spectral function $Q(\nu'/\nu'_\textrm{m})$ has been implemented correctly. The flux at frequencies above the cooling break however, shows the consequence of a finite and evolving upper cut-off $\gamma'_\textrm{M}$. A slope is reproduced that matches the prediction. It has been explained above and in EW09 how this slope now arises as a product of the interplay between the hot region and the blast wave width.

At the low frequencies, where synchrotron self-absorption plays a role, the simulations also reproduce a spectral slope that corresponds to what was expected from analytical calculations. The flux level is now dictated by the radiative transfer equation through a medium that is no longer completely transparent at these frequencies. As discussed in section \ref{ssa_section}, the resulting flux will differ by a factor of a few from \cite{Granot2002}, due to a difference in approach when calculating the absorption coefficient from the particle distribution.

We emphasize that fig \ref{spectra_figure} covers 10 orders of magnitude in frequency, 8 orders of magnitude in flux and four orders of magnitude in observer time. As we expected from analytical calculations, the spectral slopes in the different power law regimes do not change over time. The transitions between the different regimes are smooth. An explicit calculation of the sharpnesses of the transitions will be presented in a follow-up paper.

\subsection{Light curves}
\label{lightcurves_section}
\begin{figure}
\includegraphics[width=\columnwidth]{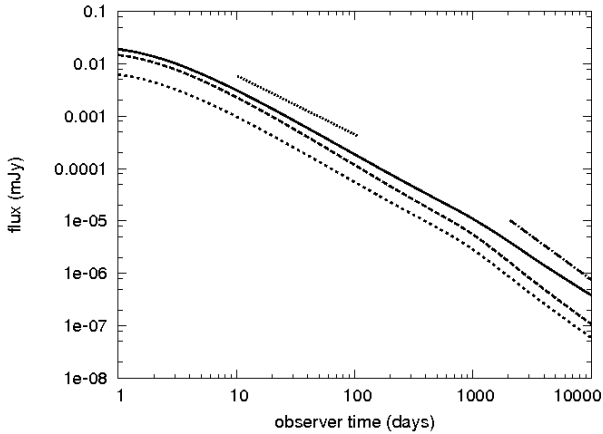}
\caption{Comparison of optical (at $t \cdot 10^{14} Hz$ light curves for different equations of state. The top curve has $\Gamma_\textrm{ad} = 4/3$, the center curve the advanced EOS and the bottom curve $\Gamma_\textrm{ad} = 5/3$. For clarity as few complications as possible are included: cooling and self-absorption are switched off, $\epsilon_B$ is fixed at $0.01$ and $\xi_\textrm{N} = 0.1$ everywhere. Also plotted are the expected relativistic slope $3(1-p)/4$ and nonrelativistic slope $(21-15p)/10$.}
\label{EOScurves_figure}
\end{figure}

We will use optical light curves to illustrate the consequences of the different assumptions and model parameters. In fig. \ref{EOScurves_figure} we present simulated light curves for simulations that differ only in the EOS used. Electron cooling and self-absorption have been disabled, $\epsilon_B$ is fixed at $0.01$ and $\xi_\textrm{N} = 0.1$ everywhere throughout the simulation. This allows for a clear view on both the effect of the EOS and of the transrelativistic break. The latter can be found at $\backsim 1000$ days for all three simulations. This is somewhat earlier than the transition time determined from the fluid flow in section \ref{blastwave_velocity_section}, which we determined to be around 1290 days (the difference is due to relativistic beaming). \emph{The transition time is also later than what is usually assumed for the nonrelativistic transition by nearly a factor three.}

\begin{figure}
\includegraphics[width=\columnwidth]{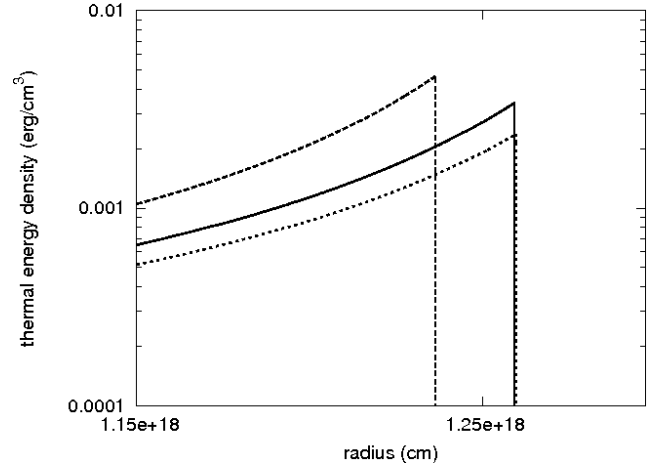}
\caption{Direct comparison between thermal energy density $e'_\textrm{th}$ profiles for the different equations of state. The top profile has $\Gamma_\textrm{ad} \equiv 4/3$, the center curve the advanced EOS and the bottom profile has $\Gamma_\textrm{ad} \equiv 5/3$. All snapshots are taken at 515 days simulation time. The difference in radius between the blast waves is 2 percent.}
\label{eprofiles_figure}
\end{figure}

The difference in flux from the different EOS assumptions can be traced to the different thermal energy profiles (and hence, for fixed $\epsilon_\textrm{B}$, to magnetic field energies that differ with the same ratios), with $\Gamma_\textrm{ad} \equiv 4/3$ having the highest $e'_\textrm{th}$. This is illustrated in fig. \ref{eprofiles_figure}. The difference in peak thermal energy densities between the fixed adiabatic index simulations is a factor of two, as expected from the ratio $(5/3-1)/(4/3-1)$. Because the flux depends on the thermal energy via the magnetic field strength and $\gamma_\textrm{m}$ (see equations \ref{power_equation} and \ref{critical_frequency_equation}), the flux for $\Gamma_\textrm{ad} \equiv 4/3$ is higher than that for $\Gamma_\textrm{ad} \equiv 5/3$. The light curve for the advanced equation of state lies between the two limiting cases, starting close to the $4/3$ curve but moving to the $5/3$ as the flow becomes nonrelativistic. \emph{This additional decrease in flux has the consequence that the advanced equation of state light curve will be slightly steeper in the transrelativistic phase than the fixed adiabatic index light curves}.

\begin{figure*}
\includegraphics[width=0.49\textwidth]{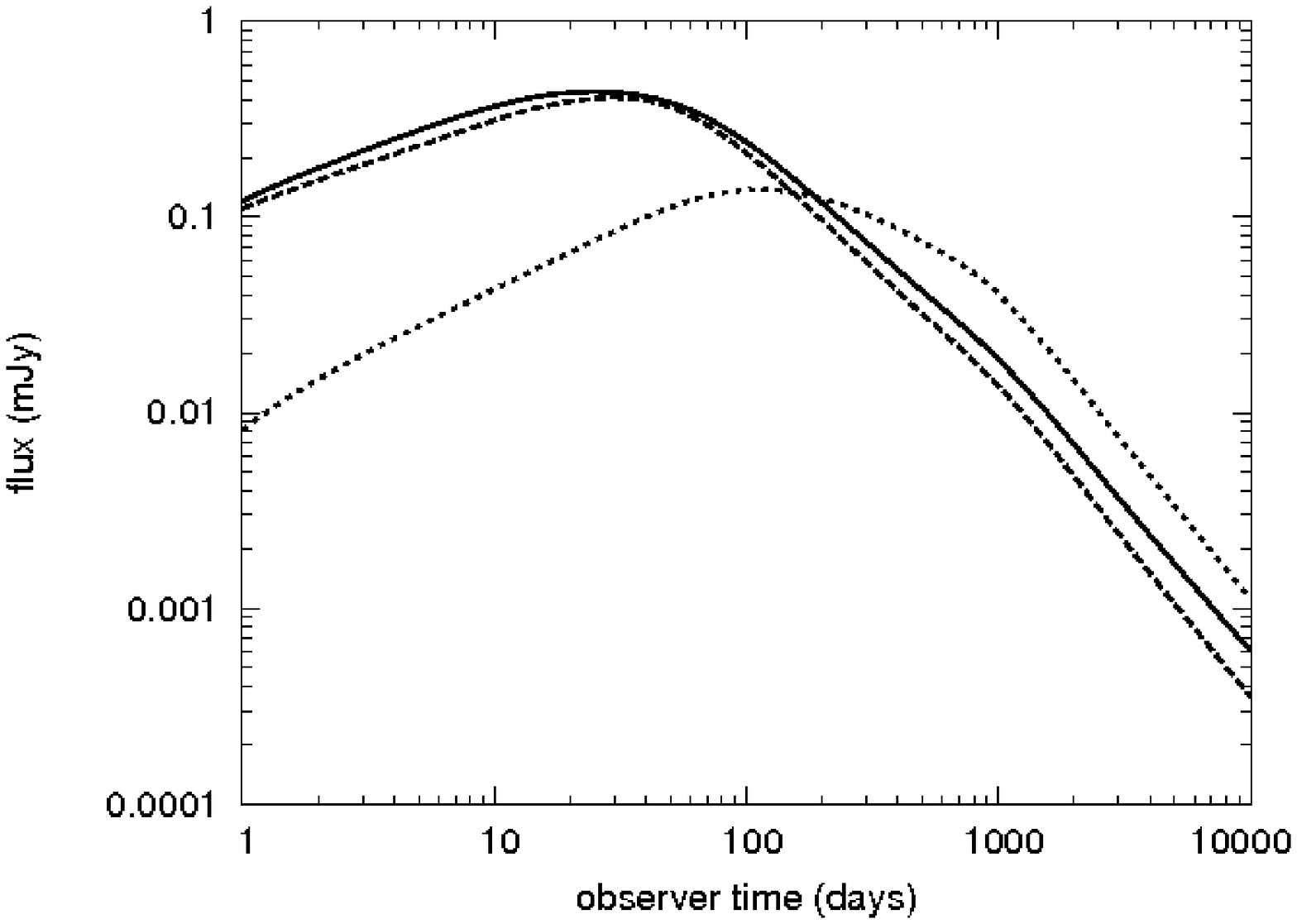}
\includegraphics[width=0.49\textwidth]{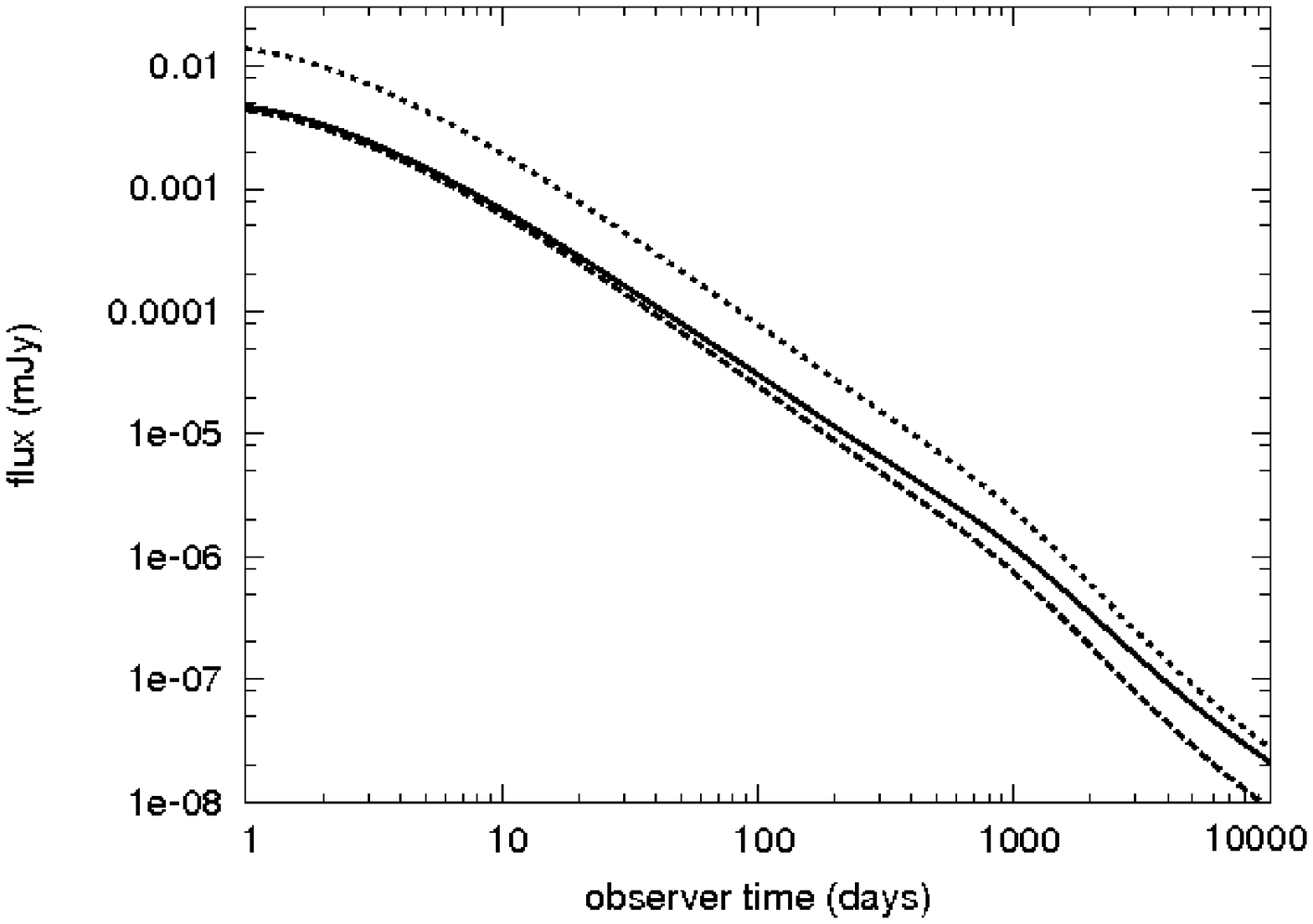}
\caption{Left: Comparison between complete simulation light curve (solid line) at radio frequency $4.8 \cdot 10^9$ Hz and simulation curves where $\xi_\textrm{N}$ is kept fixed at 1.0 (dashed line) and 0.1 (dotted line) throughout. The complete curves start out close to $\xi_\textrm{N} \equiv 1$ but slowly evolve towards $\xi_\textrm{N} \equiv 0.1$. Right: same as left, only now for optical frequency $5 \cdot 10^{14}$ Hz. As with the radio light curves, the full curve starts near $\xi_\textrm{N} \equiv 1$ but turns to $\xi_\textrm{N} \equiv 0.1$.}
\label{lightcurves_figure}
\end{figure*}
In figure \ref{lightcurves_figure} we show the effects of the detailed evolution calculation of $\xi_\textrm{N}$. Aside from the full simulations, we also perform two simulations that keep $\xi_\textrm{N}$ fixed throughout at either 1 or 0.1 but are otherwise identical to the full simulation. At early times in the radio, before the peak frequency has passed, the $\xi_\textrm{N} \equiv 1$ curve lies \emph{above} the full simulation curve, where at early times in the optical it lies below. 

\begin{figure}
\includegraphics[width=\columnwidth]{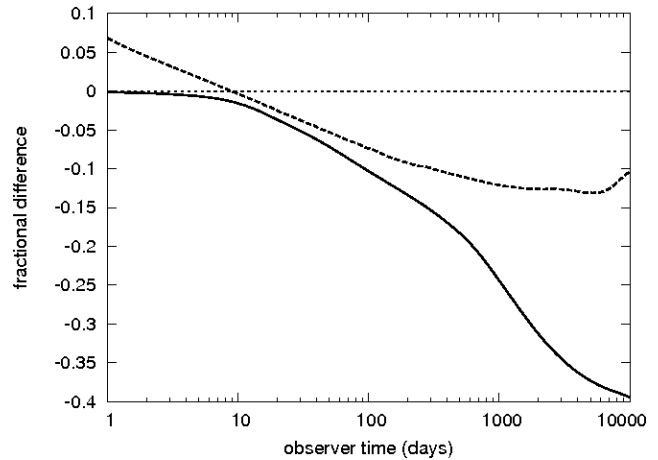}
\caption{Fractional difference between complete and fixed $\epsilon_B \equiv 0.01 $ simulation light curves. Solid line for radio, dashed line for optical.}
\label{B_figure}
\end{figure}
Figure \ref{B_figure} shows the fractional difference  between complete and fixed $\epsilon_B \equiv 0.01 $ simulation light curves (the light curves themselves lie very close to each other on a plot using a logarithmic scale), calculated via $(F_\textrm{fixed} - F_\textrm{complete}) / F_\textrm{complete}$, where $F$ is the flux. The figure shows that the late time light curves for the fixed $\epsilon_B$ end up below the light curves that trace the evolution of the magnetic field. This can be understood from the fact that evolving $e'_B$ according to $e'_B \propto (\rho')^{4/3}$ implies a relative rise of the magnetic field energy density relative to (but still far below) the thermal energy when the flow becomes nonrelativistic. Fixing $\epsilon_B$ forces the magnetic field energy density to follow $\rho^{\Gamma_\textrm{ad}}$. The flux spans many orders of magnitude over time. 

\begin{figure}
\includegraphics[width=\columnwidth]{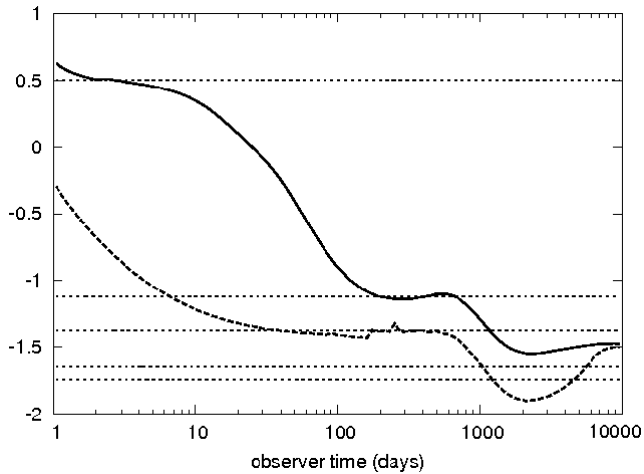}
\caption{Power law behaviour of the optical and radio light curves. The lines plot $\alpha$, assuming for every two consecutive data points the relationship $F_{i+1} = F_i ( t_{i+1} / t_i )^\alpha$. The solid line refers to the radio light curve and the dashed line to the optical light curve. The horizontal lines denote 0.5,  $3(1-p)/4 = -1.125$, $(2-3p)/4 = -1.375$, $((21-15p)/10) = -1.65$, $(4-3p)/2 = 1.75$ from top to bottom.}
\label{slopes_figure}
\end{figure}
A quantitative comparison between the slopes from the radio and optical light curves for the full simulations is shown in figure \ref{slopes_figure}. The horizontal lines in the plot indicate expected asymptotic values for the power law scalings. In the relativistic limit, the expected slope is $1/2$ before passage of $\nu_\textrm{m}$ and $-1.125$ after (using $p=2.5$). After the cooling break passes, a further steepening to $-1.375$ is expected. In the nonrelativistic regime the expected slopes before and after passage of the cooling break are $-1.65$ and $-1.75$ respectively. The plot shows that the relativistic slopes are matched very well. The radio light curve quickly tends to $1/2$ and after passage of the peak frequency it moves in $\backsim 95$ days to $-1.125$, where it remains until the onset of the nonrelativistic break time. The optical light curve starts out in the intermediate regime from the passage of $\nu_\textrm{m}$, with the passage of the cooling break coming too early for the light curve to settle into the pre-cooling break slope of $-1.125$. The post-cooling break slope $-1.375$ is obtained instead and is again maintained until the nonrelativistic break. The light curve slopes in the nonrelativistic regime are less steep than expected. A number of factors play a role here, as discussed above. The advanced EOS leads to a steepening of the decay during the transition phase (which lasts well over 10,000 days), whereas the increase in $e'_\textrm{B}$ relative to $e'_\textrm{th}$ and the decrease in $\xi_\textrm{N}$ (leading to an \emph{increase} in energy per particle) lead to less steep decay. The change in slope in the nonrelativistic regime is the result of the interplay between these different factors, with the end result being a slope less steep than expected. \emph{The final nonrelativistic slopes differ significantly from those expected from analytical models, and this has a large impact on fitting models to observational data.} 

\section{GRB030329}
\label{GRB030329_section}

In the preceding section we have systematically explored the different aspects of transrelativistic blast wave afterglows with respect to dynamics and radiation for standard values of the input parameters. We now qualitatively compare radio data for GRB030329 to simulation results using physical parameters for this GRB established by earlier authors as input. GRB030329 is one of the closest and brightest GRBs for which an afterglow was found. Because of this brightness, the afterglow could be monitored for an extended period of time at various wavelengths and after six years its radio signal is still being observed \citep{Kamble2009}. GRB030329 is a good example to use to illustrate the various aspects of the radiation code.

The redshift of GRB030329 has been determined to be $z = 0.1685$ \citep{Greiner2003}, which leads to a luminosity distance of $2.4747 \cdot 10^{27}$ cm (for a flat universe with $\Omega_M = 0.27, \Omega_\Lambda = 0.73$ and $H_0 = 71$ km s$^{-1}$ Mpc$^{-1}$). Various authors have determined the physical properties of the GRB from analytical model fits to the data (e.g. \citealt{Willingale2004, Berger2003, Sheth2003, vanderHorst2005, Huang2006, vanderHorst2008mar}) with various assumptions for the jet structure. Here we take the physical parameters established by \cite{vanderHorst2008mar}. From their conclusion for the jet break time and cooling frequency at this time, and assuming equipartition between accelerated particle energy and magnetic field energy (i.e. a fixed $\epsilon_E \equiv \epsilon_B$ in their model), we arrive at $E = 2.6 \cdot 10^{51}$ erg (for a spherical explosion), $n_0 = 0.78$ cm$^{-3}$, $p=2.1$, $\epsilon_E = \epsilon_B = 0.27$. We assume a homogeneous medium and we set the hydrogen mass fraction in this medium to unity. \citet{vanderHorst2008mar} fix $\xi_N$ at unity, but we use a nonrelativistic limit $\xi_{N,\textrm{NR}} = 0.1$. Because GRB030329 shows clear evidence of a collimated outflow, it is no longer sufficient to assume a spherical explosion. When calculating emission from a jet, we assume a hard-edged jet with opening angle 22 degrees and no lateral spreading.

\begin{figure*}
\includegraphics[width=0.3631\textwidth]{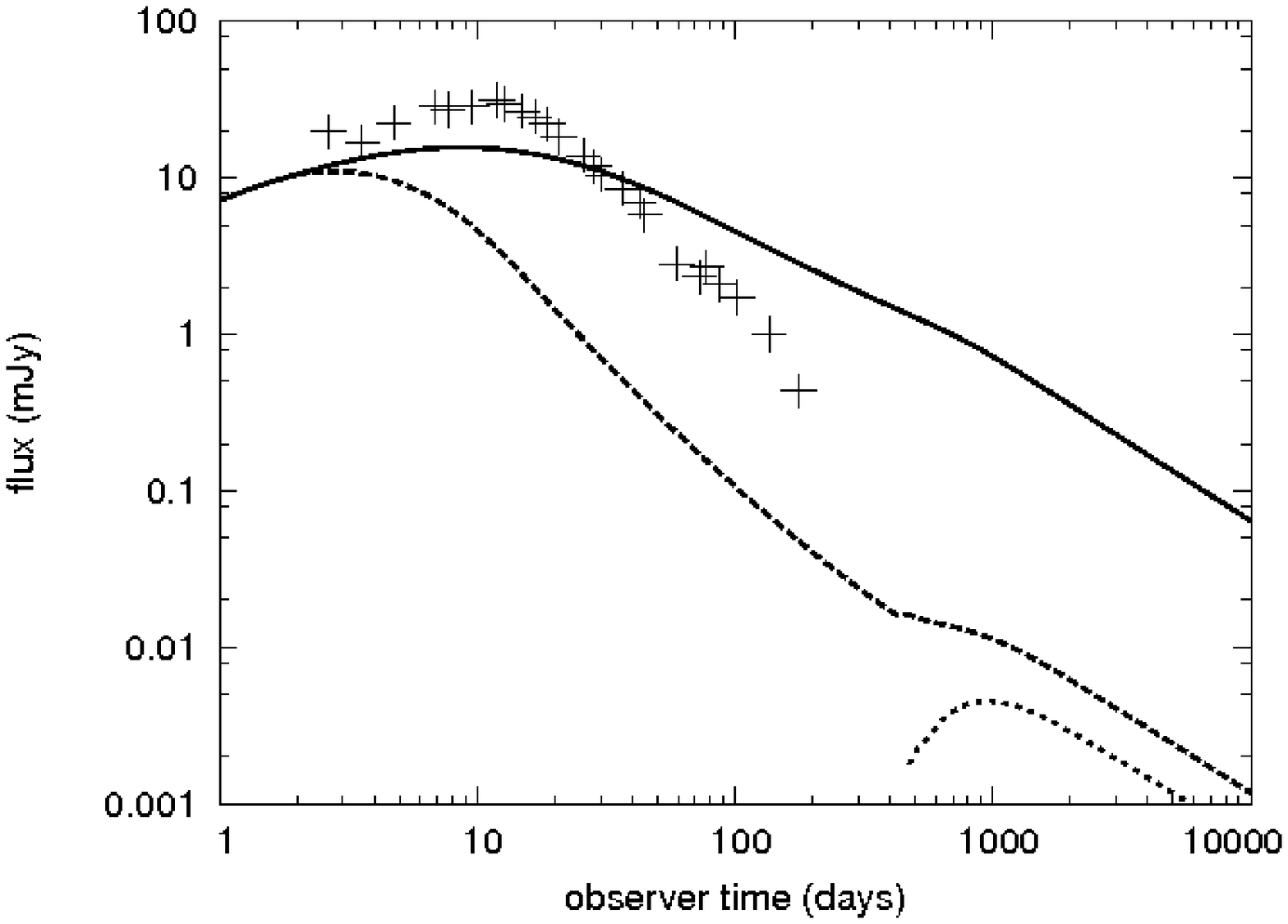}
\includegraphics[width=0.3036\textwidth]{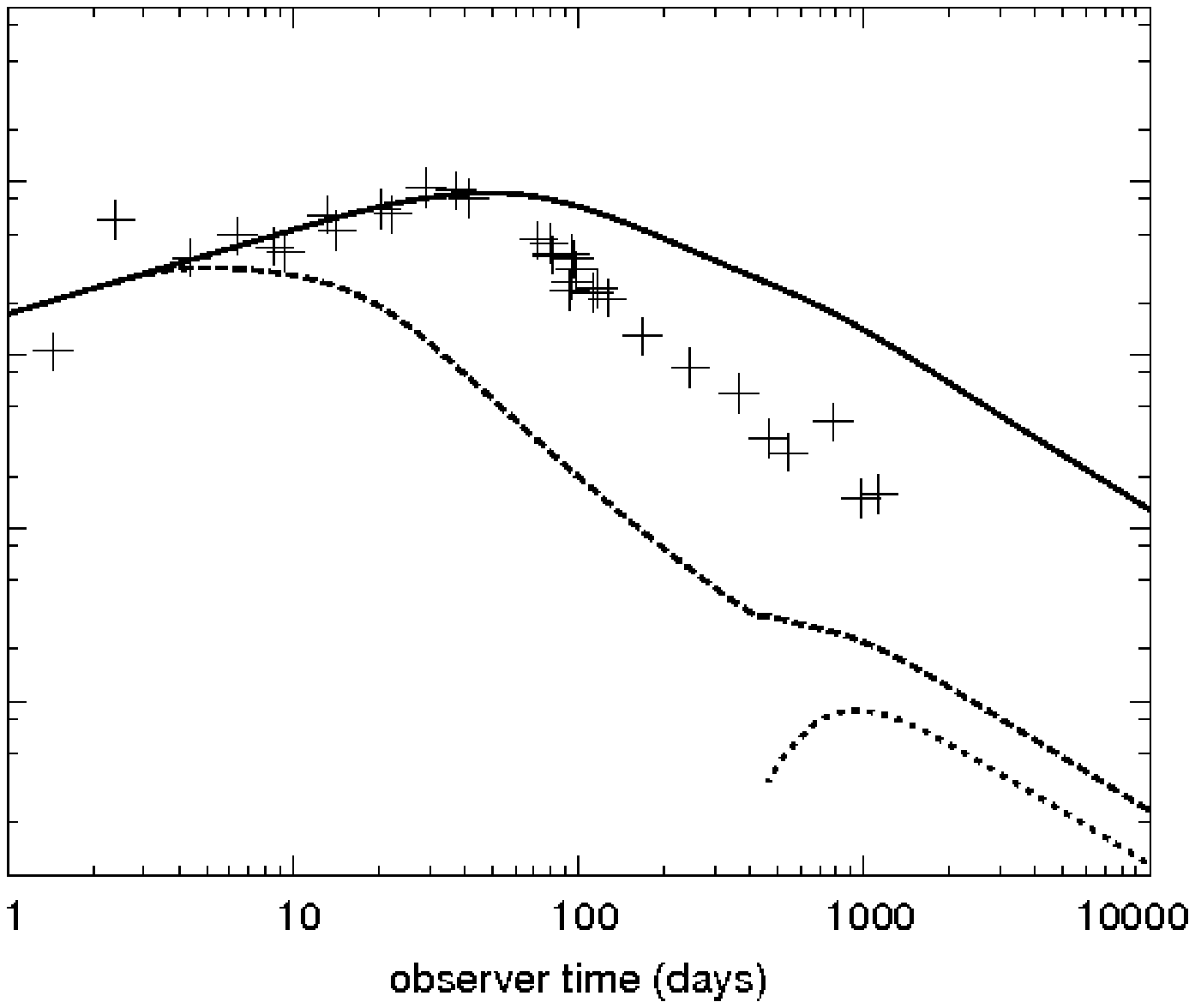}
\includegraphics[width=0.3036\textwidth]{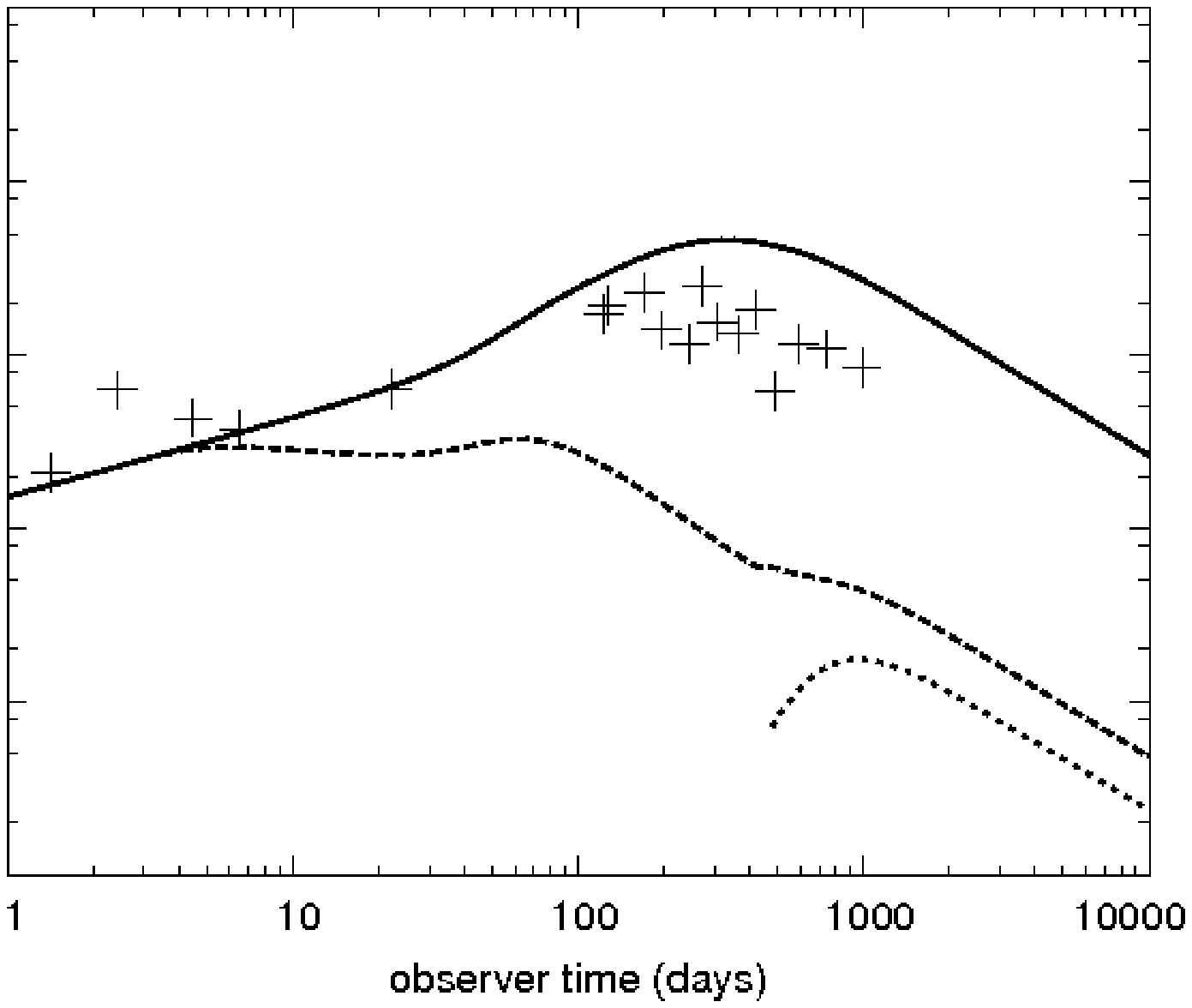}
\caption{Left: Light curves at 15 Ghz, plus data. The upper curve is calculated from a spherical explosion, the bottom from a hard-edged jet with opening angle 22 degrees, while the crosses are the data. The receding jet is clearly visible. The bottom right curve is the radiation from the counterjet alone. The transition to the nonrelativistic regime can be seen from the spherical explosion simulation at around the same time when the counterjet becomes visible. Center: Light curves at 4.8 Ghz, plus data. Right: Light curves at 1.4 Ghz, plus data.}
\label{GRB_figure}
\end{figure*}
We have plotted light curves at 15 GHz, 4.8 GHz and 1.4 GHz in figure \ref{GRB_figure}, including data points from the Westerbork Synthesis Radio Telescope (WSRT) (4.8 GHz and 1.4 GHz, \citealt{vanderHorst2005}) for comparison and the Very Large Array (VLA) (15 GHz, \citealt{Berger2003}). Two things are clearly visible. First, our simulated light curves still differ strongly from the data, although largely the same input parameters have been used for the blast wave simulations as those that were derived from fitting to the dataset using an analytical model for the blast wave. The different assumptions in \citet{vanderHorst2005} account for this in part, but nevertheless this demonstrates once more the need for detailed fit prescriptions from simulations (a similar conclusion was drawn in EW09 for the ultra-relativistic case). Second, the counterjet contribution will stand out clearly for a hard-edged jet model. For now, the comparison between simulation and data is still qualitative. Newer data are available and once the simulation input parameters are fine-tuned with respect to the data as well (as opposed to estimated using an analytical fit to the data), it should be possible to address the rise of the counterjet in a more quantitative fashion.

Because of the equipartion constraint on $\epsilon_B$ and $\epsilon_E$, both were given a relatively high value of 0.27 at the shock front. In the nonrelativistic regime, the magnetic energy density will grow relative to the thermal energy density further downstream (although both will decrease strongly in absolute value). At $t_e \backsim 34.7$ yrs, the last time covered by our simulation (set up to cover 10,000 days in observer time) we find that $\epsilon_B$ has risen to approximately $0.36$ at the back of the blast wave ($0.43$ where the blast wave density again equals the upstream density). Even further downstream, when the density has fallen three orders of magnitude below the upstream density, $\epsilon_B$ peaks at $1.28$. This is not unphysical, but merely an indication that magnetic fields have become dynamically important in a region of the fluid which has no consequence for the light curve.

\subsection{Low frequency array light curves and resolved images}
\begin{figure*}
\includegraphics[width=0.3631\textwidth]{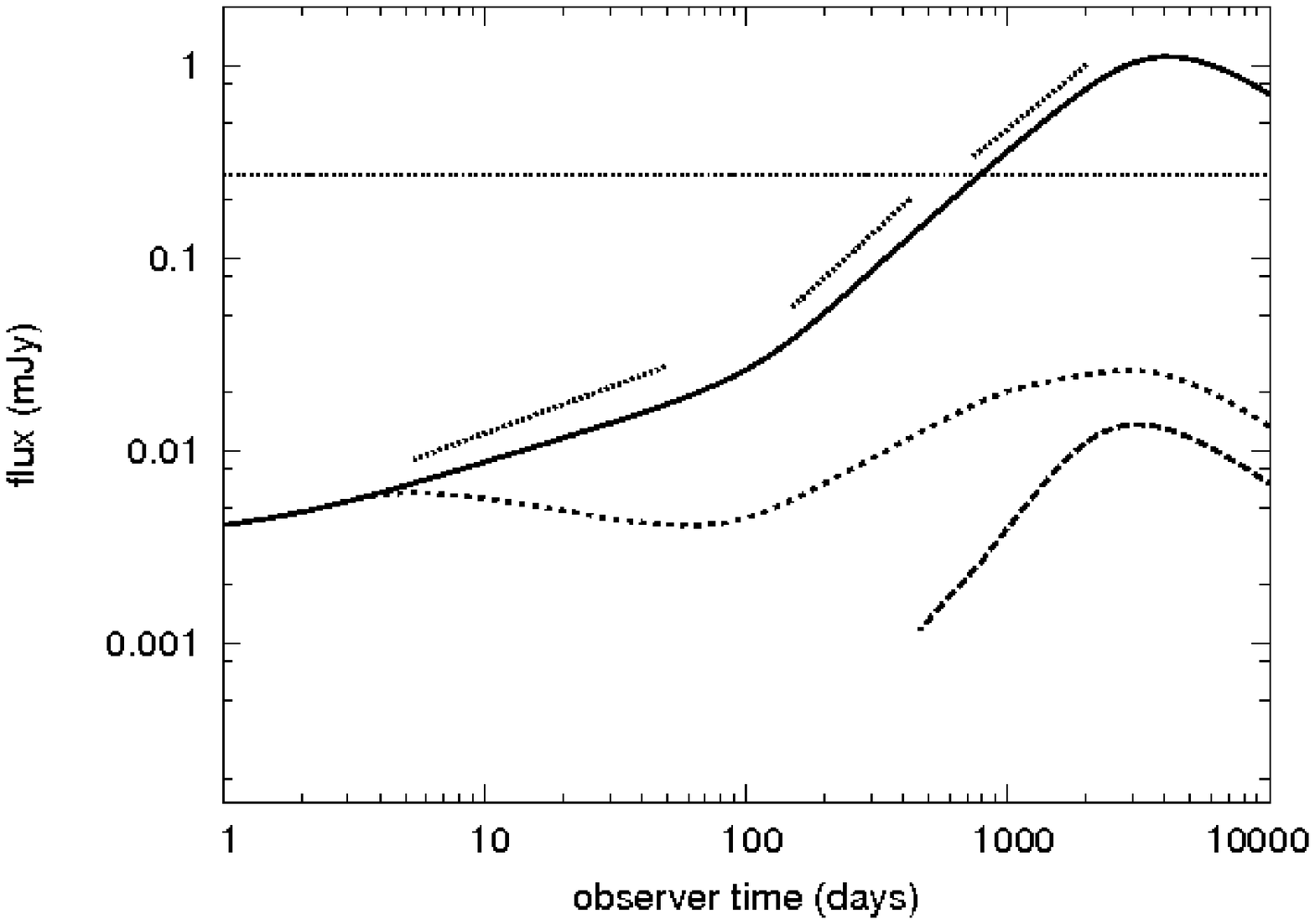}
\includegraphics[width=0.3036\textwidth]{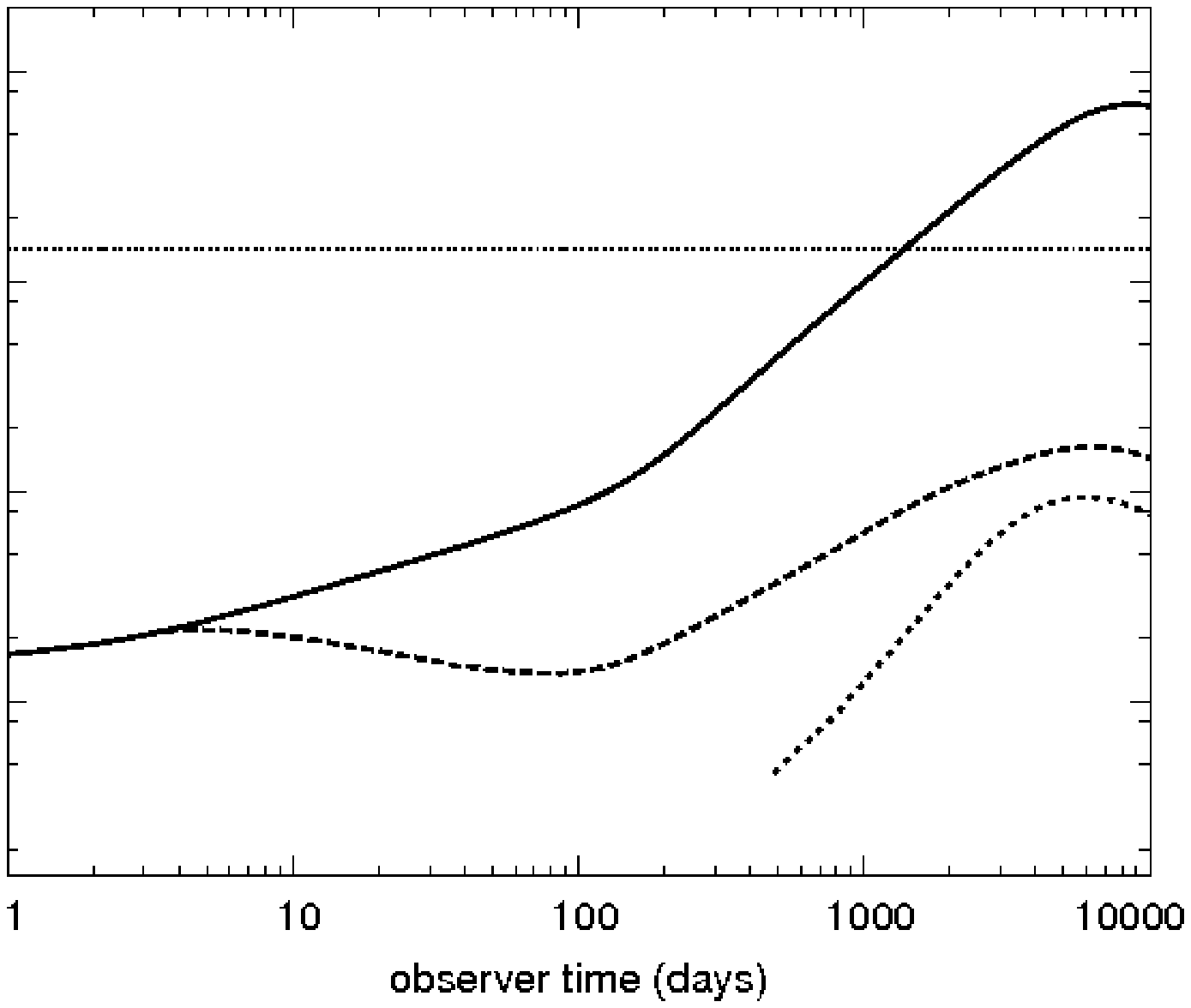}
\includegraphics[width=0.3036\textwidth]{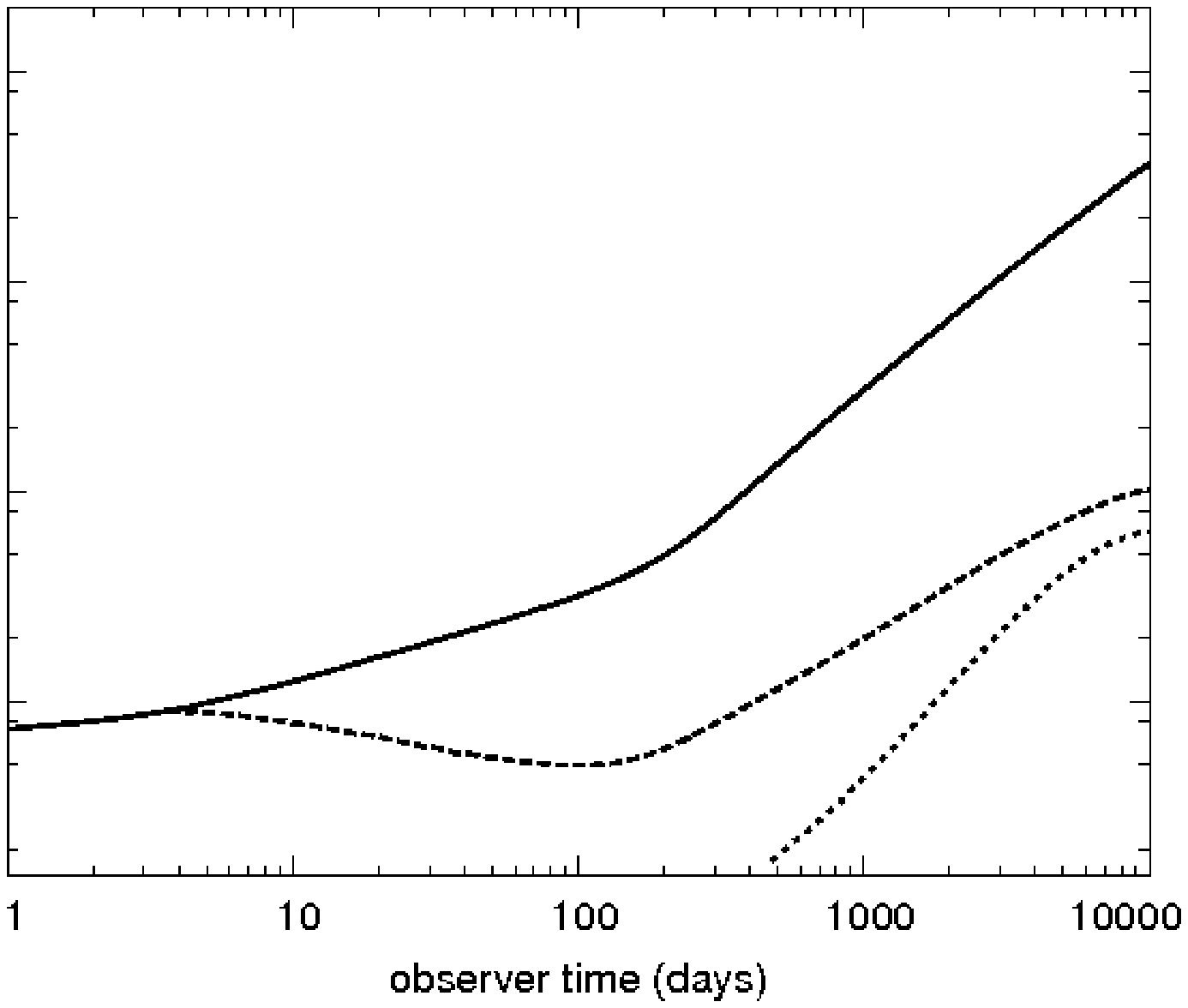}
\caption{Left: Simulated light curve at 200 MHz for GRB030329, top curve for spherical explosion and bottom curve for hard-edged jet with opening angle 22 degrees. We have drawn the following slopes from left to right: $1/2$, $5/4$ and $11/10$. LOFAR sensitivity for 25 core and 25 remote stations after four hours of integration time is 0.273 mJy and indicated by the horizontal line. Center: Simulated light curve at 120 MHz for GRB030329. LOFAR sensitivity is 0.145 mJy. Right: Simulated light curve at 75 MHz for GRB030329. LOFAR sensitivity is 4.2 mJy, too high to be shown in the plot.}
\label{lofar_figure}
\end{figure*}
Figure \ref{lofar_figure} shows predicted light curves at the very low frequencies that can be explored in the near future by radio telescopes such as the \emph{Low Frequency Array} (LOFAR), assuming four hours of integration time, 25 core stations and 25 remote stations \citep{Nijboer2009}. GRBs are among the prime targets for LOFAR's Transient Key Project \citep{Fender2006}. Most of the time all light curves lie below the self-absorption break. This, in combination with the $\nu_\textrm{m}$ break, a hard-edged jet model and the turnover to the nonrelativistic regime, leads to an interesting double peak structure of the light curve. First the signal rises, according to the relativistic rise in the self-absorption regime that predicts a slope of $1/2$. After $\backsim 4$ days a clear jet break is seen and the resulting drop in slope leads to a decreasing signal again. Around circa 150 days the critical frequency $\nu_\textrm{m}$ passes through the observed frequency band. The slope of the spherical explosion changes accordingly towards the predicted relativistic $5/4$. Around approximately 600 days the blast wave has become nonrelativistic and the counterjet starts to contribute (but is still overwhelmed by the forward jet). The predicted nonrelativistic slope for the spherical explosion is now $11/10$.

We have included LOFAR detection thresholds for four hours of integration time. These sensitivity limits are higher than those presented in \cite{vanderHorst2008mar}, because LOFAR has been scaled down in the meantime. The spherical explosion energy is an overestimation of the actual explosion energy and the flux levels corresponding to the jet simulations lie closer to what will actually be received. However, from fig. \ref{GRB_figure} it is clear that our qualitative comparison systematically underestimates the actual flux levels. Also, the integration time used in LOFAR can easily be increased, even up to days. Fig. \ref{lofar_figure} therefore does \emph{not} mean that GRB030329 will not be observable by LOFAR, but only that a larger integration time than four hours is likely required. 

\begin{figure*}
\includegraphics[width=0.32\textwidth]{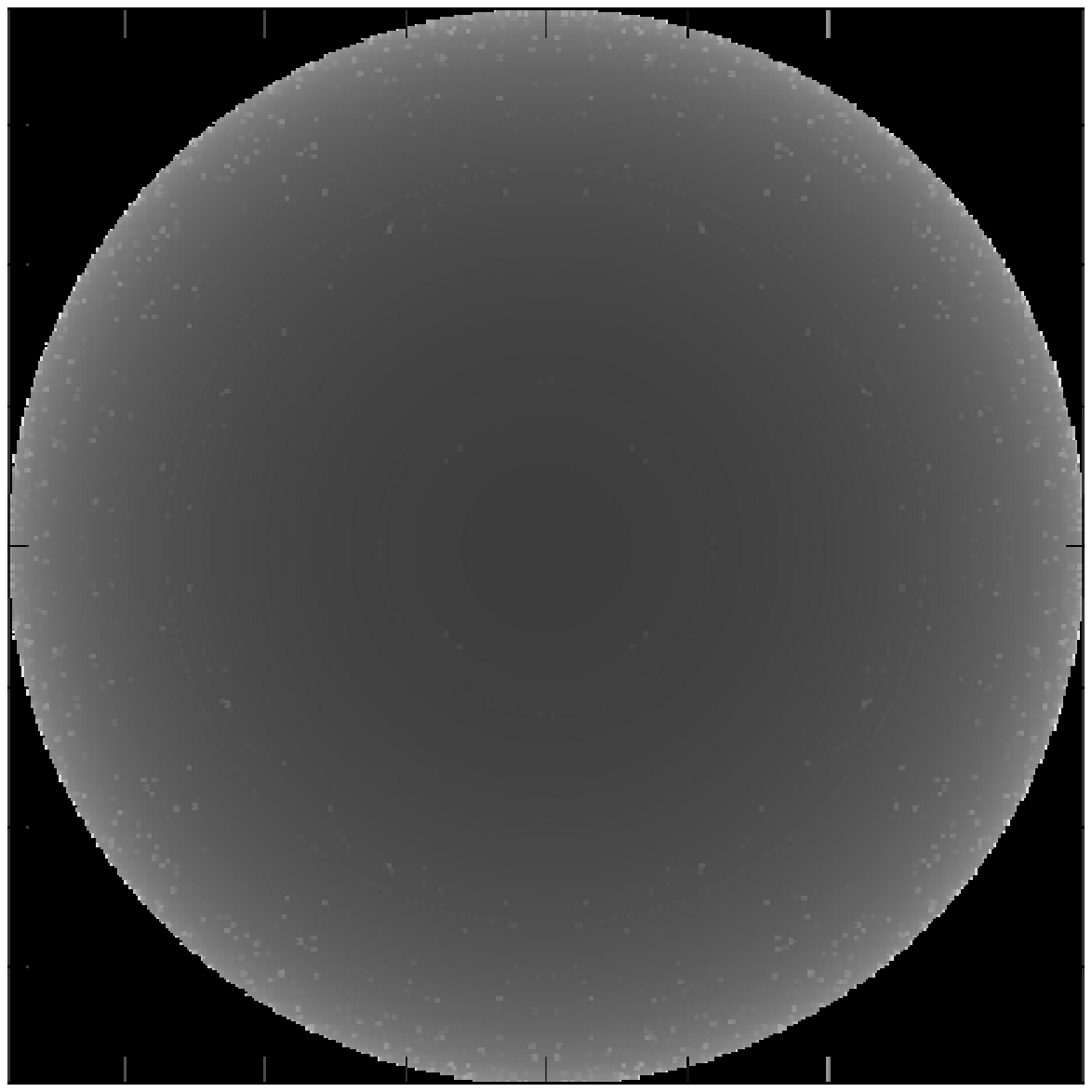}
\includegraphics[width=0.32\textwidth]{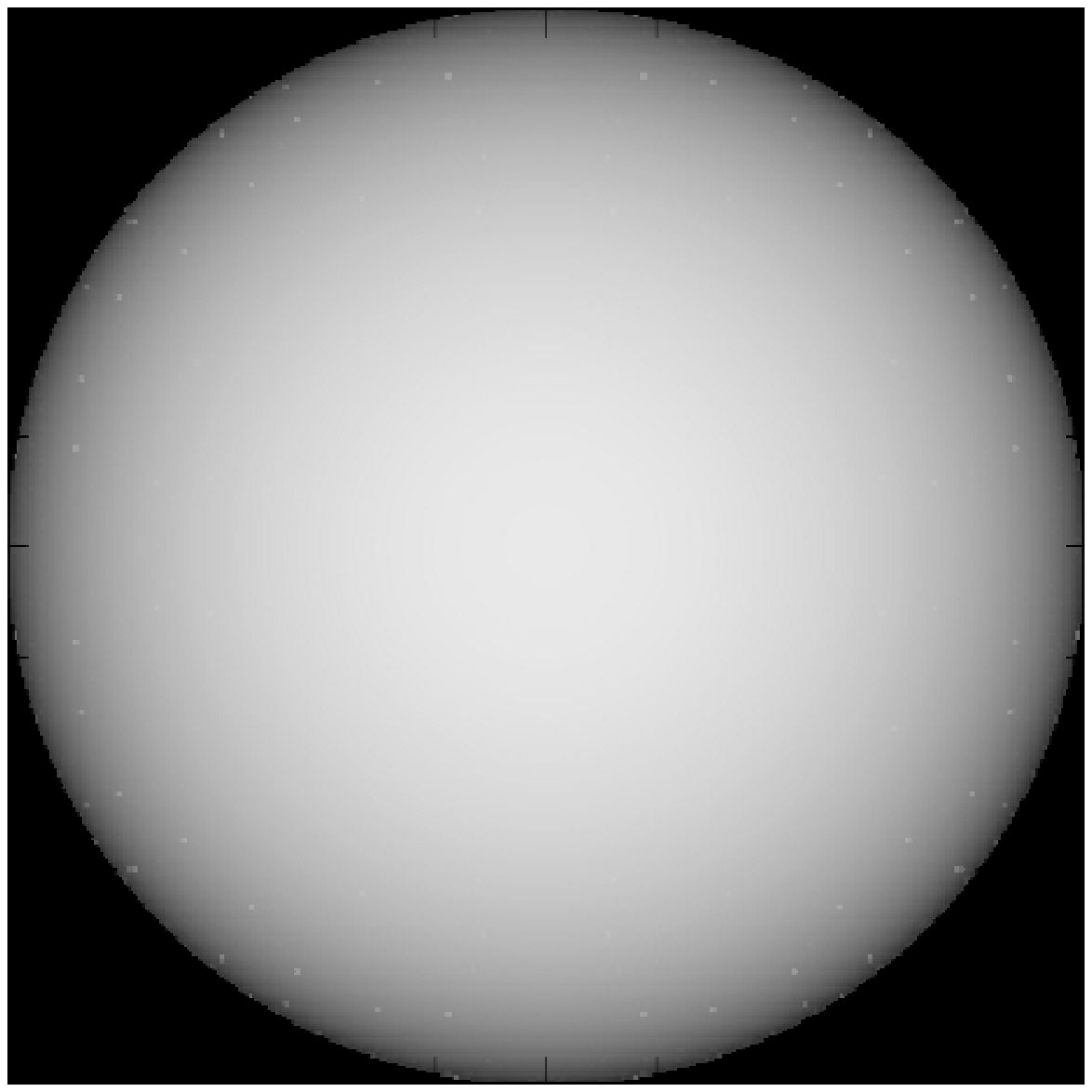}
\includegraphics[width=0.32\textwidth]{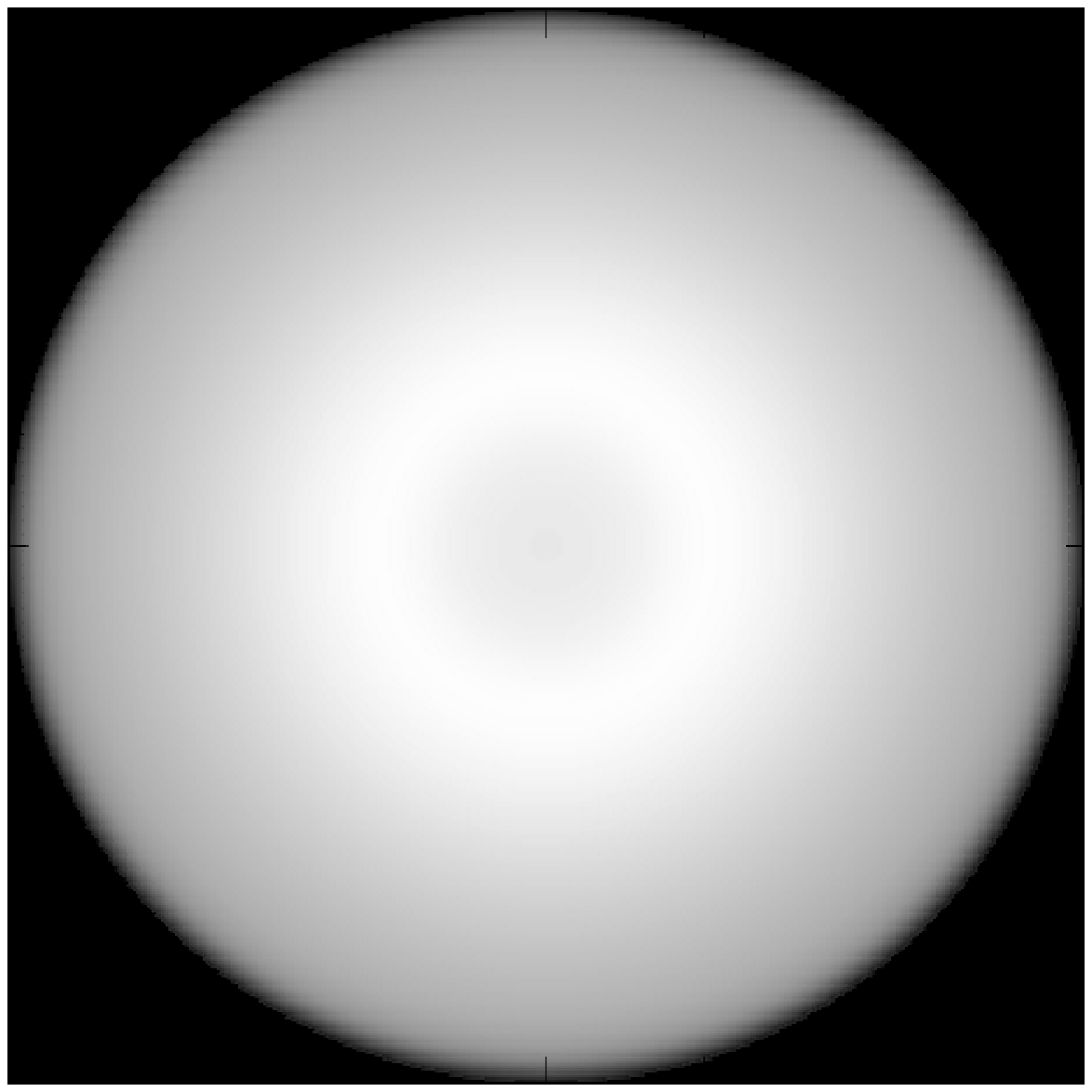}
\caption{Radio images at 200 MHz. Left: after $\backsim 16$ days. The intensity increases monotonically outward. The outer radius is $1.91 \cdot 10^{17}$ cm. Center: after $\backsim 270$ days. The intensity decreases monotonically outward. The outer radius is $9.7 \cdot 10^{17}$ cm. Right: after $\backsim 4300$ days. A central bright ring with radius $\backsim 10^{18}$ cm appears. At larger radii the intensity decreases monotonically. The outer radius is $3.4 \cdot 10^{18}$ cm.}
\label{resolved_figure}
\end{figure*}
For 200 MHz we have calculated spatially resolved images as well, for spherical explosions. Three images are presented in figure \ref{resolved_figure}, for three different observer times. They show three qualitatively different types of behaviour. At 15 days a limb-brightened image is observed, whereas at 240 days the image on the sky becomes limb-darkened. At 3900 days another structure is visible and a brighter ring exists within the image, at a radius of $\backsim 10^{18}$ cm. This is a result of the self absorption break $\nu_A$ being different for different emitting regions of the blast wave. These images are fully consistent with predictions from \citet{Granot2007} for the ultra-relativistic case.

\section{Summary and conclusions}
\label{discussion_section}
In this paper we present the results of detailed dynamical simulations of GRB afterglow blast waves decelerating from relativistic to nonrelativistic speeds, as well as spectra and light curves calculated from these simulations using a method first described in \cite{vanEerten2009} (EW09) that we have extended to include more details of synchrotron radiation. We summarize our results and conclusions below.

We have performed, for the first time, hydrodynamical simulations of decelerating relativistic blast waves using adaptive mesh refinement techniques \emph{including} a parametrisation for a shock accelerated electron distribution radiating via synchrotron radiation. From these simulated blast waves we have calculated light curves and spectra at various observer times and frequencies. An advanced equation of state was used for the dynamical simulations, with an effective adiabatic index smoothly varying between the relativistic and nonrelativistic limit. Three additional parameters were traced during hydrodynamical evolution: maximum accelerated particle Lorentz factor, magnetic field energy density and accelerated particles number density. We assumed that fewer particles were accelerated by shocks that are less relativistic. To obtain the observed flux including synchrotron self-absorption, a set of linear transfer equations were solved for beams traversing through the blast wave. This method expands upon EW09 by including self-absorption and dynamically calculated electron cooling.

We have used standard assumptions for the GRB explosion energy ($\backsim 10^{52}$ erg) and circumburst particle number density ($\backsim 1$ cm$^{-3}$) for a homogeneous medium and particle acceleration and magnetic field parameters. By directly comparing against various analytical models and expected limiting behaviour, we draw a number of conclusions about the \emph{dynamics} of our simulations: 
\begin{itemize}
 \item We find that the transition of $\beta \gamma$ directly behind the shock front from the relativistic to the nonrelativistic regime occurs later than expected, around $\backsim 1290$ days rather than $\backsim 450$ days, for the standard model parameters.
 \item An analytical calculation of $\beta \gamma$ according to \cite{Huang1999} is found to overestimate the late time values by a factor $4/3$.
 \item Directly applying the Sedov-Taylor solution to late time afterglow evolution is found to overestimate the radius by a few percent and keeping the adiabatic index fixed throughout the evolution of the blast wave will lead to systematic differences of as much as ten percent.
 \item The density jump across the shock may be arbitrarily high for relativistic shocks, but will be a factor of four in the nonrelativistic regime. This is known from the shock jump conditions. Our simulations show that the quantity $D / \gamma^2$, a combination of lab frame density and Lorentz factor directly behind the shock, will remain close to four times the unshocked density \emph{throughout the entire simulation}, even though the effective adiabatic index evolves from relativistic to nonrelativistic.
  \item If we assume the number of magnetic field lines through the surface of a fluid element a constant, the magnetic field energy will become relatively larger compared to the thermal energy. It will remain a small fraction however (assuming only a small amount of energy is used for magnetic field creation accross the shock). Our approach allows for different assumptions on the magnetic field energy evolution.
 \item The upper cut-off Lorentz factor $\gamma'_\textrm{M}$ for the shock-accelerated relativistic power law electron distribution decreases on a distance scale much smaller than the width of the blast wave due to synchrotron losses and determines the shape of the spectrum near and above the cooling break.
\end{itemize}

Using the output from the dynamical simulations, we calculate the flux. The following general conclusions are drawn for the \emph{radiation}:
\begin{itemize}
 \item Calculated light curves show a transition between the relativistic and nonrelativistic regime at around 1000 days in observer time, again later than expected.
 \item The observed fluxes for different assumptions on the equation of state may differ by a factor of a few. This is a direct consequence of the amount of thermal energy (and therefore magnetic field energy) directly behind the shock front. 
 \item Implementing a changing effective adiabatic index has the consequence that the resulting light curve will slowly evolve from the relativistic limiting value to the nonrelativistic value. This transition takes tens of years in observer time and will lead to a steeper decay in the afterglow light curve than predicted by analytical models assuming a fixed index.
 \item This steepening is a smaller effect than the combined effect of evolving the magnetic energy density and the accelerated particle number density. When all effects are included, the final light curve slopes differ markedly from the analytically expected values. This implies a significant complication for late time afterglow modeling.
\end{itemize}

We have applied our approach to GRB030329 as well, using physics parameters derived by \cite{vanderHorst2008mar} using an analytical model. It is shown that the resulting radio light curves differ up to an order of magnitude between simulation and analytical model, although this can be partly attributed to some different assumptions. Assuming a hard edged jet with an opening angle of 22 degrees, our simulated light curves show a rebrightening due to the counterjet around 1000 days. Simulated curves at radio frequencies that will be observable using LOFAR show that four hours of integration time is likely not sufficient to distinguish the signal from the noise and a larger integration time is required. Finally, spatially resolved images show a bright ring that, depending on the precise power law regime that is observed, may be located not only in the center or on the edge but also at intermediate radii within the afterglow image. This is consistent with earlier work by \cite{Granot2007} on afterglow images in the relativistic phase.

A recent paper \citep{Zhang2009} has appeared discussing afterglow blast waves decelerating to nonrelativistic velocities using twodimensional simulations. The authors find that lateral expansion of a relativistic GRB jet is a very slow process and that the jet break is mostly due to the edges of the jet becoming visible. This implies the hard edged jet model that we have applied to GRB030329 is sufficient to model the jet break at $\backsim 4$ days. \citet{Zhang2009} do not include synchrotron self-absorption and calculate the cooling break by assuming the cooling time throughout the entire blast wave equal to the grid time.

The approach to calculating light curves and spectra from generic fluid simulations that we present in this paper assumes that synchrotron radiation is the dominant radiative process, that particle acceleration takes place in a region far smaller than the blast wave width and that the feedback on the dynamics from the radiation is negligible. We briefly address these issues in appendix \ref{assumptions_section}.

\section{Acknowledgements}
This research was supported by NWO Vici grant 639.043.302 (RAMJW, KL), NOVA project 10.3.2.02 (HJvE). ZM performed computations on the K.U.Leuven High Performance computing cluster VIC, and acknowledges financial support from the FWO, grant G.0277.08, from the GOA/2009/009 and HPC Europa (project number: 228398). We wish to thank Atish Kamble for providing the radio data to GRB030329 for the qualitative comparison and Jonathan Granot for discussion on synchrotron self-absorption. 

\bibliographystyle{mn2e}
\bibliography{hveerten}

\appendix
\section{Numerical implementation}
\label{numerics_section}

\subsection{Partial differential equations}
\textsc{amrvac} was written to solve a system of coupled partial differential equations. When adding additional equations to the solver, it is therefore best to use partial differential equations. In the case of the magnetic field energy $e'_\textrm{B}$ we start by rewriting equation \ref{magnetic_field_equation} as
\begin{equation}
 \frac{ \partial}{\partial t} \frac{e'_\textrm{B}}{\rho'^{4/3}} + v^i \frac{\partial}{\partial x^i} \frac{e'_\textrm{B}}{\rho'^{4/3}} = 0.
\end{equation}
If we multiply this equation by $\rho' \gamma$ and add to this the continuity equation
\begin{equation}
 \frac{\partial}{\partial t} \rho' \gamma + \frac{\partial}{\partial x^i} \rho' \gamma v^i = 0,
\end{equation}
which we first multiply by $e'_\textrm{B} / \rho'^{4/3}$, we obtain
\begin{equation}
 \frac{\partial}{\partial t} \frac{ \gamma e'_\textrm{B}}{\rho'^{1/3}} + \frac{\partial}{\partial x^i} \frac{ \gamma e'_\textrm{B} v^i}{\rho'^{1/3}} = 0.
\label{e_B_equation}
\end{equation}
This is the type of conservation equation that \textsc{armvac} is specialized in, and it is therefore the quantity $\frac{ \gamma e'_\textrm{B}}{\rho'^{1/3}}$ that we calculate in \textsc{amrvac}.

For the evolution of the upper cut-off $\gamma'_\textrm{M}$ we follow a similar procedure. We start by simplifying equation \ref{kinetic_equation} to
\begin{equation}
\frac{\dev}{\dev t} \frac{\rho'^{1/3}}{\gamma'_\textrm{M}} = \alpha \frac{\rho'^{1/3} B'^2}{\gamma},
\label{kinetic_intermediate_equation}
\end{equation}
where $\alpha \equiv \sigma_\textrm{T} / 6 \pi m_\textrm{e} c$, and $t$ refers to lab frame time (i.e. emission time). The Lorentz factor in the source term arises when we write the comoving time derivative in the lab frame. We can now follow a procedure similar to what we did for the magnetic energy density, but first we rewrite the equation above once more for numerical reasons. The quantity $\gamma'_\textrm{M}$ varies over many orders of magnitude in a very short time span and any quantity that depends on $\gamma'_\textrm{M}$ linearly is therefore difficult to deal with numerically. We solve this by rewriting equation \ref{kinetic_intermediate_equation} into
\begin{equation}
\frac{ \dev}{\dev t} \ln \frac{ \rho'^{1/3}}{\gamma'_\textrm{M}} = \frac{ \alpha \gamma'_\textrm{M} B'^2}{\gamma}.
\end{equation}
Although there still is a linear dependence on $\gamma'_\textrm{M}$ in the source term, in practice this equation provides a better starting point for \textsc{amrvac}. From combining with the continuity equation we get
\begin{equation}
 \frac{\partial}{\partial t} \gamma \rho' \ln \frac{ \rho'^{1/3}}{\gamma'_\textrm{M}} + \frac{ \partial }{\partial x^i} v^i \gamma \rho' \ln \frac{ \rho'^{1/3}}{\gamma'_\textrm{M}} = \alpha \gamma'_\textrm{M} \rho' B'^2,
\label{gamma_M_equation}
\end{equation}
with $\gamma \rho' \ln \rho'^{1/3}/ \gamma'_\textrm{M}$ the quantity of interest. A similar approach to tracing the effect of cooling in the context of relativistic blast waves has also been taken by \citet{Downes2002}. Although in our formalism $\gamma'_\textrm{M}$ at the shock front should be reset to infinity, and therefore $\gamma \rho' \ln \rho'^{1/3} / \gamma'_\textrm{M}$ to minus infinity, we just take a very low value for $1 / \gamma'_M$ in order to minimize numerical diffusion. In our simulations, this arbitrarily low value corresponds to a hard cut-off of the spectrum above $\nu \backsim 10^{18}$ Hz, at frequencies sufficiently far above our observation range to be of no consequence. The `real' $\gamma'_\textrm{M}$ catches up with the numerical $\gamma'_\textrm{M}$ almost instantaneously.

\subsection{Shock detection method}
\label{shock_detection_appendix}
\begin{figure}
\includegraphics[width=\columnwidth]{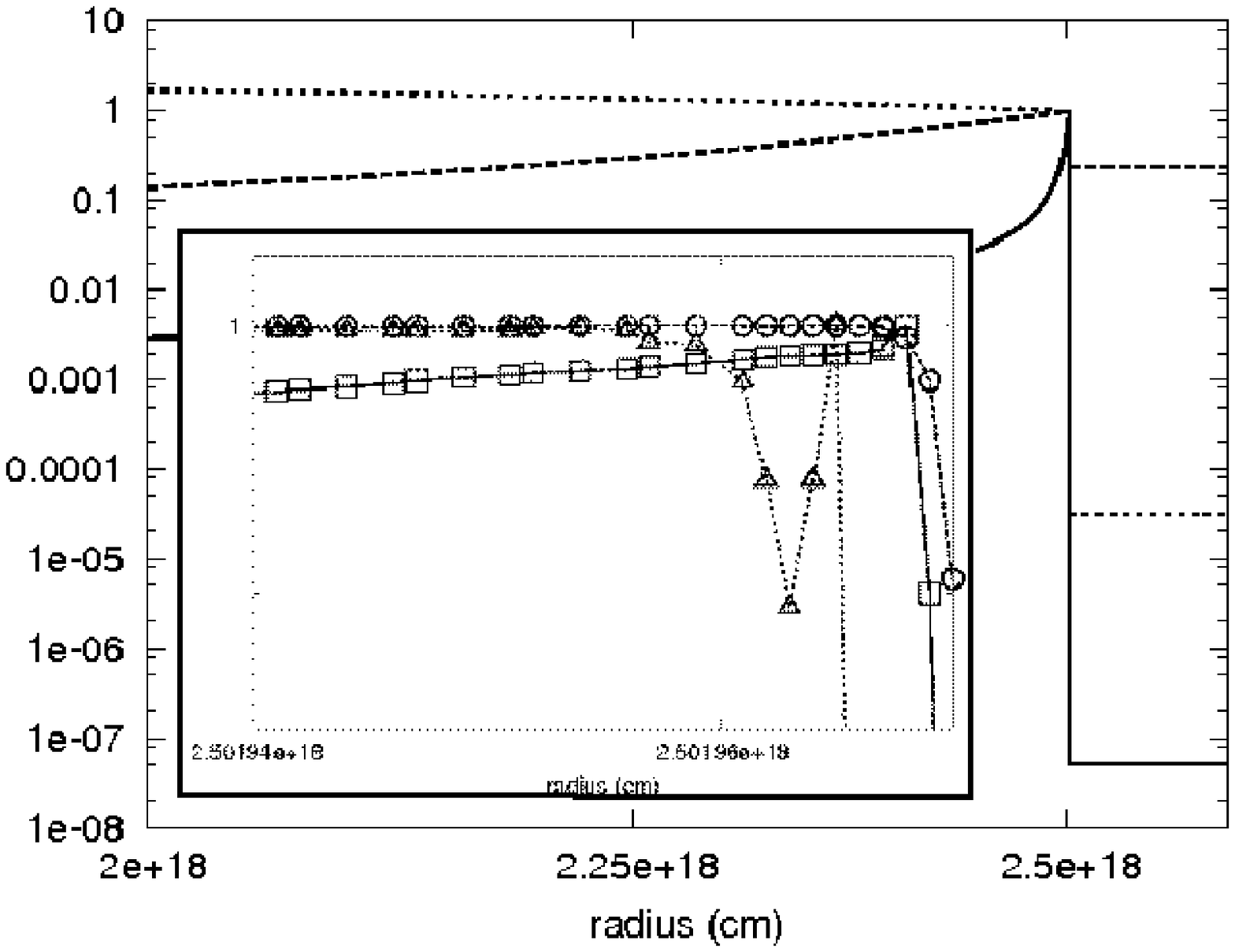}
\caption{Normalized profiles of $\gamma'_\textrm{M}$ (solid line and squares), $\rho'$ (dashed line and circles) and $\xi_\textrm{N}$ (dotted line and triangles). The actual values at the shock front are $1.9\cdot10^7$, $7.0\cdot10^{-24}$ g cm$^{-3}$ and 0.31 respectively. The simulation time is 1400 days. In the lower left corner we zoom in on the shock front, showing exactly where we reset $\gamma'_\textrm{M}$ and $\xi_\textrm{N}$.}
\label{shock_detection_figure}
\end{figure}
In total, \textsc{amrvac} now calculates the evolution of three additional quantities: $n_\textrm{acc}$ (using equation \ref{n_acc_equation}), $\gamma e'_\textrm{B} / \rho'^{1/3}$ (using equation \ref{e_B_equation}) and $\gamma \rho' \ln ( \rho'^{1/3} / \gamma'_\textrm{M} )$ (using equation \ref{gamma_M_equation}). All three quantities get reset wherever a shock is detected. Both the reset values of $n'_\textrm{acc}$ and $e'_\textrm{B}$ depend on the fluid variables directly behind the shock front and it is therefore important that we determine the position of the shock front as accurately as possible. Mathematically speaking, a shock is a discontinuity in the flow variables with a sudden increase in entropy across the discontinuity. In practice, however, finding a shock in a numerical approximation is more involved, both due to numerical shock diffusion and because, strictly speaking, there is a shock discontinuity across every grid cell boundary. 

This has the consequence that if we try to find shocks by checking for discontinuities or for entropy jumps, we will find both shocks all over the numerical diffused shock region and at a random variety of positions where the numerical noise happens to rise above a predetermined shock treshold. This then implies that we keep on resetting the additional quantities over some region, something which is especially unwanted in the case of $\gamma'_\textrm{M}$, given our approach where we take a fluid cell to contain a collection of electrons that have been shocked exactly at the same time and we critically rely on the size of the \emph{hot region} (see section \ref{electron_cooling_section} and EW09, appendix D, for details).

Because the shocked particle number density and the magnetic field density depend directly on the fluid variables, using, for example, a jump in $\beta \gamma$ as a trigger, as has been done by \citet{Downes2002}, is not an option in these cases either. Although it serves as an excellent indicator of the front of a shock, it will not point us to a location where we can find information on the strength of the shock, but to an arbitrarily defined position just in front of that. 

In this paper we solve the issues of shock detection with two shock detection algorithms, both of them making use of the fact that $D/\gamma^2$ directly behind the shock is four times the density just in front of the shock. For $n'_\textrm{acc}$ and $e'_B$ we define the shock front to be at the peak of the Lorentz factor profile, in the region where $D / \gamma^2 > 3.5 \rho_0$. The numerical constant is arbitrary and could be taken closer to 4. With this method we ensure that the shock is detected at those positions where the fluid quantities are sufficiently close to their peak values, although multiple shock peaks may be detected in close proximity of each other due to numerical noise.

For $\gamma'_\textrm{M}$ it is essential that we only detect a single shock front. Here we care less about the precise fluid variable values. For the purpose of resetting $\gamma'_\textrm{M}$ we define the shock front to be at that position where $D / \rho_0 \gamma^2$ \emph{crosses} the value $3.5$. For a single shock front, this only happens once. Although, in principle, $\gamma \beta$ could have been used instead of $D / \rho_0 \gamma^2$, the latter offers the significant advantage that it does not change in scale over the course of the simulation and always remains close to 4, whereas $\gamma \beta$ becomes arbitrarily small.

Figure \ref{shock_detection_figure} illustrates the use of the Lorentz factor profile peak as a shock detector. It shows that the numerical diffusion is really very small and that $\gamma'_M$ changes over a significantly smaller spatial scale than $\rho'$.

\subsection{Synchrotron self-absorption}
\label{ssa_subsection}

Equation \ref{ssa_equation} can also be expressed as
\begin{eqnarray}
 {\alpha}_{\nu'}' & = & K\,{\nu '}^{-2}\,\int\limits_{{\gamma}_{m}'}^{{\gamma}_{M}'}\dev\gamma_{e}' \,\mathcal{P}\big(\frac{\nu '}{\nu_{cr,e}'}\big) \times \nonumber \\
 & & \Big[ (p+2){\gamma_{e}'}^{-(p+1)}\big(1-\frac{\gamma_{e}'}{\gamma_{M}'}\big)^{p-2}\, + \nonumber \\
& &  \, (p-2) \frac{{\gamma_{e}'}^{-p}}{{\gamma_{M}'}} \big (1 - \frac{\gamma_{e}'}{\gamma_{M}'} \big)^{p-3} \Big],
\end{eqnarray}
where
\begin{equation}
K = C\,\frac{\sqrt{3} q_e^3 B'}{8\pi m_e^2 c^2}.
\end{equation}
Here we have used the fact that $N_e(\gamma'_e) \propto \gamma'_e ( 1 - \gamma'_e / \gamma'_M)$ (i.e. a slightly modified powerlaw distribution). The scaling factor ($C$) of $N_e(\gamma'_e)$ is determined in terms of $\gamma'_M$ and $\gamma'_m$ from the requirement that the total number of accelerated electrons constitutes a fixed fraction $\xi_N$ of the available electrons. The symbol $\mathcal{P}$ denotes the pitch angle (the angle between magnetic field and particle velocity) averaged version of the synchrotron function, a dimensionless function representing the shape of the synchrotron spectrum for a single electron in the same way $\mathcal{Q}$ represents the spectrum of a distribution of particles. The $\nu_{cr,e}'$ in the argument is connected to $\gamma'_e$ via equation \ref{critical_frequency_equation} (see also EW09).

By changing variables from ${\gamma '}_{e}$ to $y=\dfrac{\nu '}{{\nu '}_{cr,e}}$ we obtain:
\begin{eqnarray}
 {\alpha}_{\nu'}' & = & \frac{K}{2}\,\Big(\frac{4\pi m_e c}{3q_e B'}\Big)^{-\frac{p}{2}}\,{\nu'}^{-\frac{p+4}{2}} \times \nonumber \\
& & \big[I_1(y_{M},y_{m}) \, + \, I_2(y_{M},y_{m})\big],
\end{eqnarray}
where the quantities $I_1(y_{M},y_{m})$ and $I_2(y_{M},y_{m})$ are:
\begin{equation}
 I_1 \equiv (p+2) \, \int\limits_{y_{M}}^{y_{m}} \dev y\,\mathcal{P}(y)\,y^{\frac{p-2}{2}}\,\big( 1-(\frac{y_M}{y})^{1/2} \big)^{p-2}
\end{equation}
and
\begin{equation}
 I_2 \equiv (p-2) \,y_M^{1/2}\,\int\limits_{y_{M}}^{y_{m}} \dev y \, \mathcal{P}(y)\,y^{\frac{p-3}{2}}\,\big( 1-(\frac{y_M}{y})^{1/2} \big)^{p-3}.
\end{equation}

As in the case of $\mathcal{Q}(y_M,y_m)$ (see \citealt{vanEerten2009}) values of $I_1(y_{M},y_{m})$ and $I_2(y_{M},y_{m})$ are tabulated for moderate $y_M$ and $y_m$, whereas their limiting behavior, for extreme values of $y_M$ and $y_m$ is analytically estimated. Namely, if $y_M/y_m\rightarrow 1$ the integrals of both $I_1$ and $I_2$ reduce to the expression inside the integral, evaluated at $y_m$, multiplied by the appropriate range in $y$-space, i.e. $(y_m-y_M)$.

For $y_M\ll 1$ the integrals' behaviour becomes hard to analytically estimate, especially for general values of $y_m$ and $p$. Instead, we fit approximate expressions to the values extrapolated from the tables.

In the case that $\frac{y_m}{y_M}\gg 1$ and $y_m$ is outside the tabulated values, we can break the integral into two parts by using the last tabulated value $\tilde{y}_{m}$. For $I_2$ the formula is:

\begin{eqnarray}
 I_2 & = & (p-2)\,y_M^{1/2} \times \nonumber \\ 
 & & \Bigg[\, \int\limits_{y_M}^{\tilde{y}_{m}} \dev y\,\mathcal{P}(y)\,y^{\frac{p-3}{2}}\,\big( 1 - (\frac{y_M}{y})^{1/2} \big)^{p-3} + \nonumber \\
& & \mathcal{Q}(y_m)y_m^{\frac{p-1}{2}}\,-\,\mathcal{Q}(\tilde{y}_{m}){\tilde{y}_{m}}^{\frac{p-1}{2}} \Bigg] .
\end{eqnarray}

For $I_1$ the result is identical, only the terms inside the square brackets (including $\mathcal{Q}(x)$) have to be evaluated for $p \to p+1$.

Finally, for $y_M \gg 1$ the result of both integrals is approximated by zero.

\subsection{Adaptive mesh and linear radiative transfer}
\label{adaptive_mesh_subsection}
\begin{figure}
\includegraphics[width=\columnwidth]{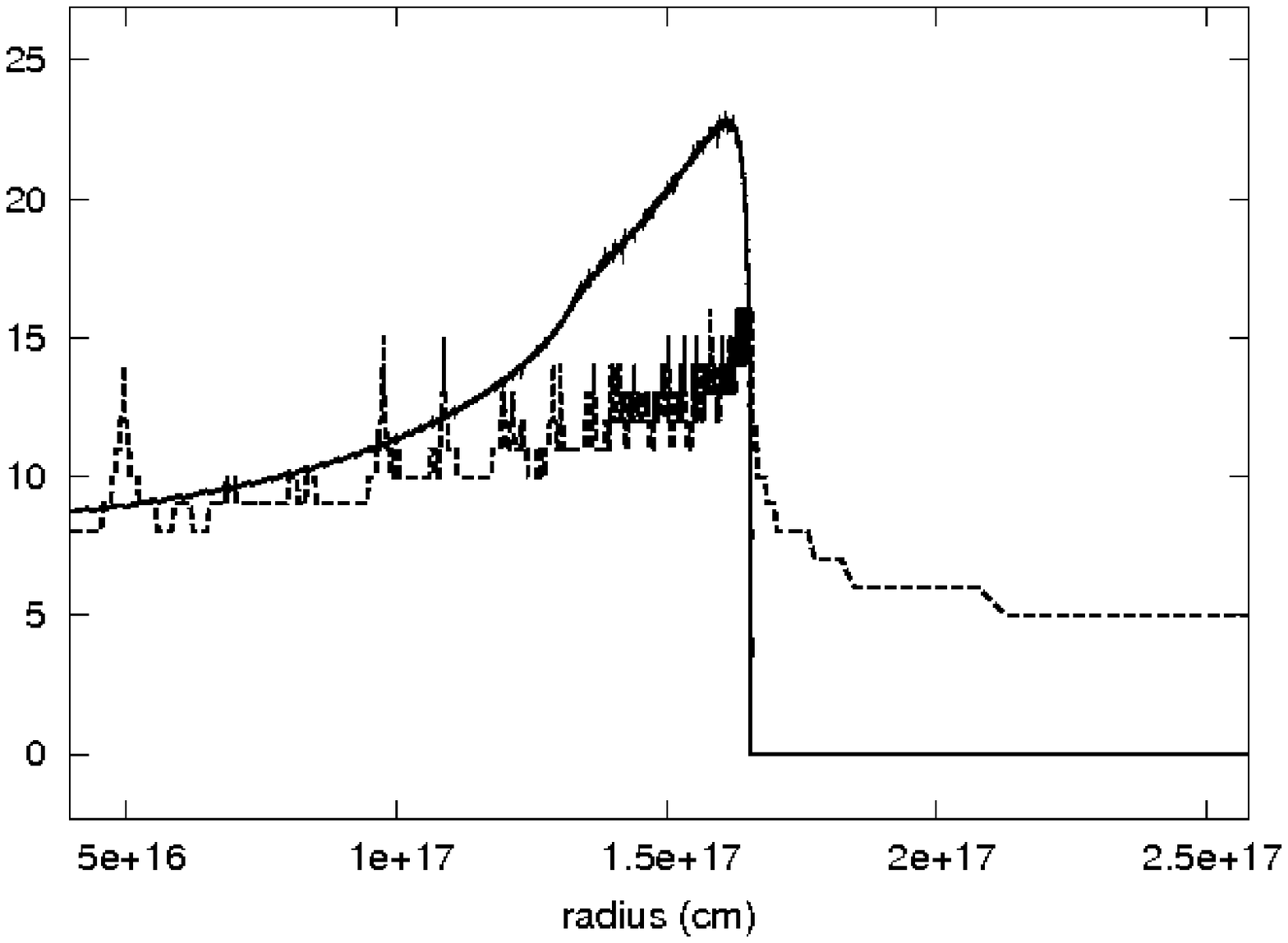}
\caption{Intensity and refinement levels perpendicular to the axis between the observer and the source. The maximum refinement level drops quickly to zero away from the edge of the jet. The intensity has been rescaled to an arbitrary scale suitable for direct comparison between intensity profile and refinement levels. Note that the lowest refinement level is zero.}
\label{amr_figure}
\end{figure}
We do not integrate over $A$ in equation \ref{flux_equation} directly, but resolve the different rays instead. After the integral over $t_\textrm{e}$ is finished (i.e. the bundle of linear radiative transfer equations is solved) we can integrate over $A$ to obtain the flux, while the unintegrated result provides a resolved picture of the emission from the fluid. For spherically symmetric fluid flow or for an observer positioned along the symmetry axis of a jet, the intensity on surface $A$ is symmetric around a central point. Because the fluid itself moves at nearly the speed of light, it is not a priori clear how many rays need to be included and how they should be spaced along $A$ in order to obtain a good resolution. An efficient response to this dilemma is to apply the adaptive-mesh refinement concept to $A$. The equidistant surface (EDS) $A$ contains a grid with every grid cell containing the intermediate results for a single ray. Every four neighbouring cells in each direction on the EDS are grouped together in a single block. The EDS area $dA$ that each cell represents may differ, and if the resolution threatens to become too low to adequately capture the radiation profile, a block will be split in half along each direction, spawning new blocks that represent half the size of the parent block along each direction. The refinement criterion that is used is that the combined flux from a given block must not differ by more than 1 percent (or a lower treshold, as set by the user) from the combined flux from a coarsened version of the block where only the odd cells are taken into account (with the odd cells representing an appropiately increased surface element). Neighbouring blocks may differ one refinement level at most. We have plotted an example of this strategy in figure \ref{amr_figure}. In practice we set the maximum refinement level similar to that of the fluid simulation. We also use the fluid simulation grid refinement structure to determine the  starting refinement structure of the EDS at each iteration for the transfer equation solver, in order to make sure that we will also capture the blast wave when it still has a small radius.

\section{Applicability of our model}
\label{assumptions_section}

The radiation code is written to be generally applicable to output from relativistic fluid dynamics simulations. However, a number of assumptions and simplifications have been made that are dependent on the physical context. In this appendix we briefly discuss the consequences and relevance of our assumptions in the case of GRB afterglow blast waves decelerating down to nonrelativistic speeds. We discuss the relevance of an alternative radiative process, inverse Compton scattering, of our assumption that particle acceleration takes places in a region much smaller than the blast wave width and of adiabatic expansion of the blast wave with the radiation losses having no effect on the dynamics.

\subsection{Importance of inverse Compton scattering}
A limitation to the applicability of our approach arises from the fact that inverse Compton (IC) radiation, which is not calculated, becomes important when the ratio $P'_{\textrm{syn}}/P'_{\textrm{IC}}$ approaches, or drops below unity. This ratio is also equal to the ratio between the corresponding energy fields that power the emission $e'_{\textrm{B}}/e'_{\textrm{ph}}$ (\citealt{Rybicki1986}), with $e'_{\textrm{ph}}$ being the energy density of the (synchrotron) radiation field. The effect of IC emission on the emitted spectra has been thoroughly investigated in \cite{Sari2001}. In this paper we focus only on its influence on cooling rates, as the high-energy synchrotron spectrum is expected to dominate IC emission for a wide range of physical parameters and radii.

Instead, however, of calculating the entire photon energy density due to synchrotron radiation we can use the fact that the cross-section for IC scattering drops fast beyond the Thomson limit (\citealt{Blumenthal1970}). Thus, we can define an `effective' photon field for an electron of Lorentz factor $\gamma'_e$ as
\begin{equation}
e'_{\textrm{ph,eff}}(\gamma'_{\textrm{e}})=\frac{4\pi}{c}\: \int\limits_{0}^{\nu'_{\textrm{Thom}}}I'_{\nu',\textrm{syn}}\, d\nu' ,
\end{equation}
where $\nu'_{\textrm{Thom}}=\frac{m_{\textrm{e}}c^2}{\gamma'_{\textrm{e}} h}$ (with $h$ denoting Planck's constant) is the photon frequency for which the scattering occurs marginally within the Thomson regime for a head-on collision and $I'_{\nu',\textrm{syn}}$ is the synchrotron specific intensity. Approximating the specific intensity by
\begin{equation}
 I'_{\nu',\textrm{syn}} \backsim \xi_N n' B' \left( \frac{\nu'}{\nu'_m} \right)^{(1-p)/2} R / \Gamma,
\end{equation}
and employing the analytical relations of the BM solution we find that right behind the shock front
\begin{eqnarray}
\frac{P'_{\textrm{syn}}}{P'_{\textrm{IC}}} & \approx & 7.5 \cdot 10^{16}\,f(p)\, (5\cdot 10^{-8})^{\frac{3-p}{2}}\,\xi_{\textrm{N}}^{p-2}\,\epsilon_{\textrm{B}}^{\frac{3-p}{4}}\times \nonumber \\
& & \epsilon_{\textrm{E}}^{1-p}\, n_0^{-\frac{p+1}{4}}\, (\gamma'_{\textrm{e}})^{\frac{3-p}{2}} \, R^{-1} \Gamma^{\frac{5-3p}{2}},
\end{eqnarray}
where $f(p)=(p-1)^{p-2}\,(p-2)^{1-p}\,(3-p)$, $\Gamma$ is the Lorentz factor of the shock front and $R$ the shock radius. By plugging in standard values of this paper ($\xi_{\textrm{N}}=1,\ p=2.5,\ n_0=1$ cm$^{-3},\ \epsilon_B=10^{-2},\ \epsilon_E=10^{-1},\ E=10^{52}\, \text{erg}$) and making further use of the BM equations we find for $\gamma'_{\textrm{m}}$
\begin{equation}
\frac{P'_{\textrm{syn}}}{P'_{\textrm{IC}}} \approx 1.8\cdot 10^{-10}\, R^{0.5} .
\label{Psyn/Pcompton_equation}
\end{equation}
This means that IC will dominate synchrotron energy losses for the lowest energy electrons throughout the relativistic phase of the fluid. A comparison of IC to adiabatic cooling, using the synchrotron loss term from eq. \ref{kinetic_equation} and $(\dev \gamma'_m / \dev t)_{\textrm{syn}} / (\dev \gamma'_m / \dev t)_{\textrm{IC}} = P'_{\textrm{syn}} / P'_\textrm{IC}$, gives
\begin{equation}
\frac{\big( d\!\gamma'_{\textrm{m}}/dt \big)_{\textrm{ad}} }{\big( d\!\gamma'_{\textrm{m}}/dt \big)_{\textrm{IC}} }= 10^{-43}\,R^{2.5}.
\end{equation}
The corresponding radius after which adiabatic expansion will sharply take over is about $1.6\cdot 10^{17}\, \textrm{cm}$. Moreover, for an electron of energy $\gamma'_{\textrm{e}}=10^4\, \gamma'_{\textrm{m}}$ (i.e. on the order of $\gamma'_M$), synchrotron losses will prevail at approximately $3\cdot 10^{17}\, \textrm{cm}$. Therefore it is only early on, and certainly not close to the subrelativistic transition, that IC cooling will affect the evolution of $\gamma'_{\textrm{M}}$ or $\gamma'_{\textrm{m}}$ for the assumptions made in this paper.

\subsection{Gyral radius}
\label{gyral_radius_appendix}
The gyroradius for an electron with Lorentz factor $\gamma_\textrm{e}$ is given by $r'_\textrm{g} = \gamma_\textrm{e} m_e c^2 / q_\textrm{e} B'$. Using the BM solution and eq. \ref{critical_frequency_equation} we find that for the most energetic electrons at the shock front that contribute to received flux within the frequency range under consideration ($10^8 - 10^{18}$ Hz) this radius lies below
\begin{equation}
r_{\textrm{g}}=5.6\cdot 10^{-75}\,\nu_{\textrm{cut-off}}^{1/2}\,R^{33/8}\,\textrm{cm},
\end{equation}
where $\nu_{\textrm{cut-off}}$ is the cut-off frequency in the lab frame, used to set $\gamma'_{\textrm{M}}$ at the shock front (i.e. a frequency safely above $10^{18}$ Hz). This should be compared against the size of the hot region, a measure of the spatial distance over which electrons cool significantly. The cooling \emph{time} for high energy electrons that cool on a scale much shorter than the scale over which the fluid variables change, is approximately equal to $t_\textrm{cool} \approx 6 \pi m_e c / \sigma_\textrm{T} (B')^2 \gamma'_\textrm{M}$ when the electron cools down to $\gamma'_\textrm{M}$. Using the self-similar parameter as an intermediate step, this can be linked to a spatial size of the hot region for $\gamma'_{\textrm{M}}$:
\begin{equation}
\delta_{\textrm{hot}}=8\cdot10^{-52}\,\nu_{\textrm{cut-off}}^{-1/2}\,R^{33/8}\,\textrm{cm}.
\end{equation}
Thus, for $\nu_{\textrm{cut-off}}=10^{18-21}\,\textrm{Hz}$ the gyroradius of the most energetic electrons is many orders of magnitude smaller than the size of the corresponding hot region, justifying our assumption that particle acceleration takes place locally near the shock front (which we have implemented by a local injection of hot electrons) and the use of an advection equation to model the evolution of $\gamma_M$.

\subsection{Feedback on the dynamics}
A last issue is the possibility of the radiative energy losses becoming comparable to the initial energy load of the fireball ($E$) that would imply a considerable impact on the dynamics of the flow. This could be explicitly quantified by calculating the total radiative output during the simulations and comparing it to the explosion energy. However, we can address this issue in a more qualitative manner by noting that the low energy electrons cool predominantly by causing the expansion of the volume they are occupying (slow cooling), even at the shock front. Moreover, for values of $p>2$ (which is the case under consideration) these electrons are the main energy carriers. In combination with the fact that the total energy residing in relativistic electrons is limited by $\epsilon_{\textrm{E}}$ (typically of the order of $10\%$), we are confident that the total energy radiated through synchrotron, especially in the subrelativistic regime, will be orders of magnitude smaller than $E$, and thus not affect considerably the dynamics.

\end{document}